\documentclass[%
 reprint,
 amsmath,amssymb,
 aps,
pra,
]{revtex4-2}
\usepackage{amsmath}
\usepackage{amssymb}
\usepackage{bbold}
\usepackage{bm}
\usepackage{braket}
\usepackage{caption}
\usepackage{csquotes}
\usepackage{derivative}
\usepackage{dcolumn}
\usepackage{dsfont}
\usepackage{float}
\usepackage{graphicx}    
\usepackage{lineno}
\usepackage{mathtools}
\usepackage{physics}
\usepackage{slashed}
\usepackage{subcaption}  
\usepackage{tikz}
\usepackage{verbatim}
\usepackage{xcolor}

\usepackage{ragged2e} 
\usepackage[percent]{overpic}

\usetikzlibrary{decorations.pathmorphing}
\usetikzlibrary{calc}
\usepackage{hyperref}

\usepackage{xr} 
\usepackage{booktabs}
\usepackage{nicematrix} 
\definecolor{giallo}{RGB}{255, 255, 0}
\definecolor{rosso}{RGB}{255, 0, 0}
\definecolor{verde}{RGB}{0, 128, 0}
\definecolor{blu}{RGB}{0, 0, 255}
\definecolor{white}{RGB}{255, 255, 255}

\begin{document}

\preprint{APS/123-QED}


\title{Local ergotropy dynamically witnesses many-body localized phases}

\author{F. Formicola$^{1,*,\dag}$}\author{G. Di Bello$^{1,*,\dag}$}\author{G. De Filippis$^{2,3}$}\author{V. Cataudella$^{2,3}$}\author{D. Farina$^{1}$}\author{C. A. Perroni$^{2,3}$} 
\affiliation{$^{1}$Dip. di Fisica E. Pancini - Università di Napoli Federico II - I-80126 Napoli, Italy}
\affiliation{$^{2}$SPIN-CNR and Dip. di Fisica E. Pancini - Università di Napoli Federico II - I-80126 Napoli, Italy}
\affiliation{$^{3}$INFN, Sezione di Napoli - Complesso Universitario di Monte S. Angelo - I-80126 Napoli, Italy}
\affiliation{$^*$Corresponding authors: F. Formicola, francesco.formicola@unina.it and G. Di Bello, grazia.dibello@unina.it}
\affiliation{$^{\dag}$Both authors contributed equally.}



\begin{abstract}
Many-body localization is a dynamical phenomenon characteristic of strongly interacting and disordered many-body quantum systems which fail to achieve thermal equilibrium. From a quantum information perspective, the fingerprint of this phenomenon is the logarithmic growth of the entanglement entropy over time. We perform intensive numerical simulations, applied to a paradigmatic model system, showing that the local ergotropy, the maximum extractable work via local unitary operations on a small subsystem in the presence of Hamiltonian coupling, dynamically witnesses the change from ergodic to localized phases. Within the many-body localized phase, both the local ergotropy and its quantum fluctuations slowly vary over time with a characteristic logarithmic law analogous to the behaviour of entanglement entropy. This showcases how directly leveraging local control, instead of local observables or entropies analyzed in previous works, provides a thermodynamic marker of localization phenomena based on the locally extractable work.

\end{abstract}

\maketitle


Numerous studies have investigated dynamical localization phenomena, in which quantum many-body systems fail to reach thermodynamic equilibrium \cite{Nandki,Abanin,Alet,Sierant}. Many-body localization (MBL), characteristic of strongly interacting and disordered many-body systems, is qualitatively different from localization of non-interacting systems. The hallmark of MBL is the development of non-local correlations, which, however, lead only to a logarithmic growth of entanglement entropy over time \cite{serbyn2013universal,vedi_parametri,lukin2019probing}. Since directly measuring the entanglement entropy is notoriously challenging (see \cite{lukin2019probing}), it is fundamental to use alternative experimental markers \cite{schreiber2015observation,smith2016many,Choi,Kohlert,gong2021experimental} to characterize MBL which still has many elusive aspects \cite{Sierant}.  

The properties of MBL can be relevant not only for fundamental research, but also for applications. One promising field of application is that of 
quantum batteries 
\cite{campaioli2024colloquium,le2018spin,ferraro2018high}. Recent works have focused not only on the charging process and energy storage \cite{andolina2018charger,farina2019,mitchison2021charging,
dou2022cavity,yao2022optimal, mazzoncini2023optimal}, but also on work extraction \cite{andolina2019,arjmandi2022performance}, where the main investigated quantity is the ergotropy \cite{joshi2022experimental, hu2022optimal}, defined as the maximum amount of work extractable from an isolated quantum system via cyclic unitary transformations \cite{allahverdyan, alicki2013entanglement}. Assuming $\hat{H}$ to be the system Hamiltonian and $\hat{\rho}$ the statistical operator describing the quantum state, the ergotropy $ \mathcal{E}$ is a functional of $\hat{\rho}$ and $\hat{H}$:  
\begin{equation}
\begin{split}
    \mathcal{E}\hspace{0.5mm}&[\hat{\rho}, \hat{H}] \coloneqq 
    \max\limits_{\hat{U} \in \hspace{0.5mm}\mathcal{U}(d)}  \Tr{\hat{H}\bigg(\hat{\rho}-\hat{U}\hspace{0.5mm}\hat{\rho}\hspace{0.8mm}\hat{U}^{\dag}\bigg)}.
\end{split}
    \label{global_erg}
\end{equation}
The maximization has to be carried out over the set $\mathcal{U}$ of all available unitary operators $\hat{U}$ acting on the system $d$-dimensional Hilbert space $\mathcal{H}$, yielding a quantity which does not vary over time for an isolated system \cite{campaioli2024colloquium,allahverdyan,lenard1978thermodynamical,pusz1978passive}.

It was shown in Ref.\,\cite{Polini} that the dynamical behaviour of the ergotropy of subsets of quantum cells of a quantum spin chain distinguishes Anderson localization (AL), MBL and ergodic (ERG) phases. The low level of entanglement in MBL phase ensures significantly superior work extraction capabilities compared to ERG phase. 
However, the protocol used in Ref.\,\cite{Polini} requires an energy cost for switching on/off the interaction between the subsystems, and, as a consequence, a form of non-local control. Moreover, a clear signature distinguishing AL and MBL phases comes only from the comparison of their temporal energy fluctuations \cite{Polini}.


The main aim of the present work is to show that a dynamical signature of localization phenomena is provided by a recently defined quantity, the local ergotropy, the maximum work extractable by local unitary operations, i.e., without turning off any Hamiltonian coupling \cite{SalviaGiovannetti}.
We analyse the one-dimensional disordered XXZ Heisenberg model, which is one of the most commonly used to investigate localization effects \cite{Abanin,Sierant}, and which is potentially realizable in laboratory \cite{murmann2015antiferromagnetic}. 
Our results are obtained using numerical techniques based on matrix product state (MPS) representation of quantum many-body states and operators \cite{perez2006matrix, white1992density,schollwock2011density}. We employ the time dependent variational principle (TDVP) to compute the time evolution \cite{haegeman2011time,haegeman2016unifying}, and a Bayesian optimization algorithm (BOA) \cite{frazier2018tutorial} to evaluate the maximum extractable work via local operations.

We investigate the dynamics of several quantities, focusing on the local ergotropy and its quantum fluctuations for a small subsystem within the spin chain. 
The subsystem can be interpreted as an open quantum battery experiencing interaction with its surrounding (the rest of the chain) under non-weak coupling conditions.
This differs from alternative perspectives where full chains are considered as quantum batteries \cite{grazi2024prldimer} and generally require forms of nonlocal control to extract work.

Over time, we have found that, in the ERG case, the local ergotropy rapidly drops to zero, in the AL phase, it stabilizes to a constant value, and, in the MBL phase, it slowly decreases following a characteristic logarithmic law. Moreover, the quantum fluctuations exhibit a similar logarithmic time behaviour which, like for entanglement entropy, turns out to be a robust feature. 

{\it Model and numerical methods.}---
We consider a one-dimensional disordered XXZ Heisenberg model with nearest-neighbour interactions between $1/2$ spins. 
The Hamiltonian reads as
\begin{equation}
     \begin{split}
         \hat{H}=&\frac{J_{\perp}}{2} \sum\limits_{i=1}^{N-1} \bigg\{\hat{S}_{i}^{+}\hat{S}_{i+1}^{-}+\hat{S}_{i}^{-}\hat{S}_{i+1}^{+} \bigg\}
     \\&+ J_z\sum\limits_{i=1}^{N-1} \hat{S}_{i}^{z}\hat{S}_{i+1}^{z}+\sum\limits_{i=1}^N h_i\hspace{1mm}\hat{S}_i^{z},
     \end{split}
     \label{H_spin}
\end{equation}
where $\hbar=1$. For each chain site $i$, we have the Pauli operator $\hat{S}_i^{z}=\frac{1}{2}\hat{\sigma}_i^{z}=\frac{1}{2}\big(\ket{\uparrow}\bra{\uparrow}_i-\ket{\downarrow}\bra{\downarrow}_i\big)$ and the ladder operators $\hat{S}_{i}^{-}=\ket{\downarrow}\bra{\uparrow}_i$ and $\hat{S}_{i}^{+}=\hat{S}_{i}^{-}{}^\dagger$.
$J_{\perp}$ is the spin-flip transverse coupling energy, $J_z$ is the longitudinal coupling energy, and $h_i$ are on-site random energy values, related to an effective magnetic field, extracted by a flat probability distribution of width $2W$, such that
\begin{equation}
    h_i \in \big[ -W, W \big ].
\end{equation}
Hard-wall boundary conditions are imposed for simplicity on a chain with size $N$ in the subspace $S^{z}_{tot}=0$, being $S_{tot}^{z}=\sum_{i=1}^N S_i^z$ the z-component of the total spin. 
This model can be mapped to a one-dimensional disordered spinless electron model with nearest-neighbour particle-particle interactions, through the Jordan-Wigner transformation \cite{suppmat}.
We remark that this is one of the most investigated models for localization phenomena, as it exhibits ERG, AL and MBL phases \cite{Abanin,Alet,Sierant}. In the following, we will use $J_{\perp}$ as reference energy scale.

Being the full compound SE always in a pure state its ergotropy\,\eqref{global_erg} (later on referred as \textit{global} ergotropy) coincides with its excitation energy (constant in time) \cite{suppmat}. 
However, having a unitary operator acting on the whole Hilbert space means that an \textit{experimenter} has full control over the whole isolated system. This is sometimes an unrealistic assumption. 

We shall now consider a different scenario, in which the isolated quantum system is partitioned into two subsystems $S$ and $E$. Hence, the total Hamiltonian $\hat{H}$ reads:
\begin{equation}
    \hat{H} = \hat{H}_S\otimes\hat{I}_E + \hat{I}_S \otimes \hat{H}_E + \hat{V}_{SE},
\end{equation}
where $\hat{H}_S$ and $\hat{H}_E$ are the Hamiltonians for the subsystems $S$ and $E$, respectively, while $\hat{V}_{SE}$ is the interaction operator.
In the following, we will assume that the experimenter has control only over the subsystem $S$, having the $d_s$-dimensional Hilbert space $\mathcal{H}_{S}$. This leads us to consider $S$ as an open system, in contact with $E$, which plays the role of an environment. The system $SE$ is initially in a high energy state (a charged state of the potential battery) which is not an eigenstate of the Hamiltonian $\hat{H}$, such that $\hat{\rho}(t)$ is the resulting time-evolving statistical operator. AL and MBL are dynamical phenomena, where non-stationary quantities are used as markers for localization \cite{Abanin,Sierant}.

\begin{figure}[!htbp]
  \begin{overpic}[width=1\columnwidth]{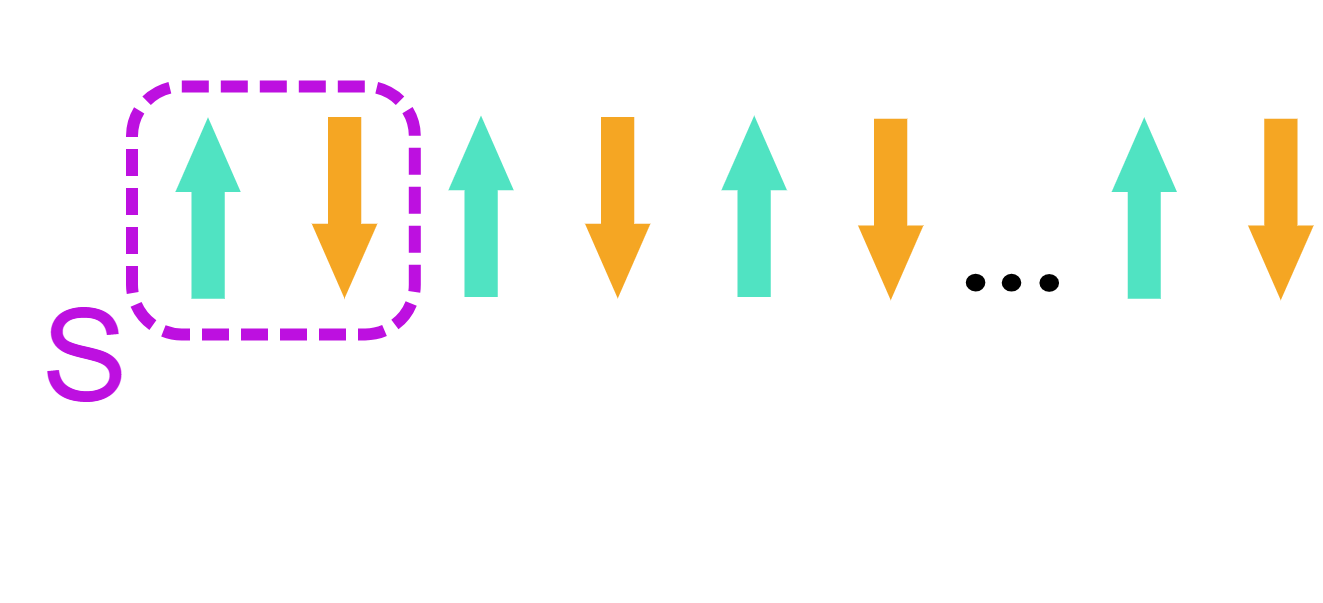}
   \put(-2.4, 37.6){\scalebox{1.5}{\textbf{(a)}}}  
  \end{overpic}
  \hspace{1.7cm}  
  \begin{overpic}[width=1\columnwidth]{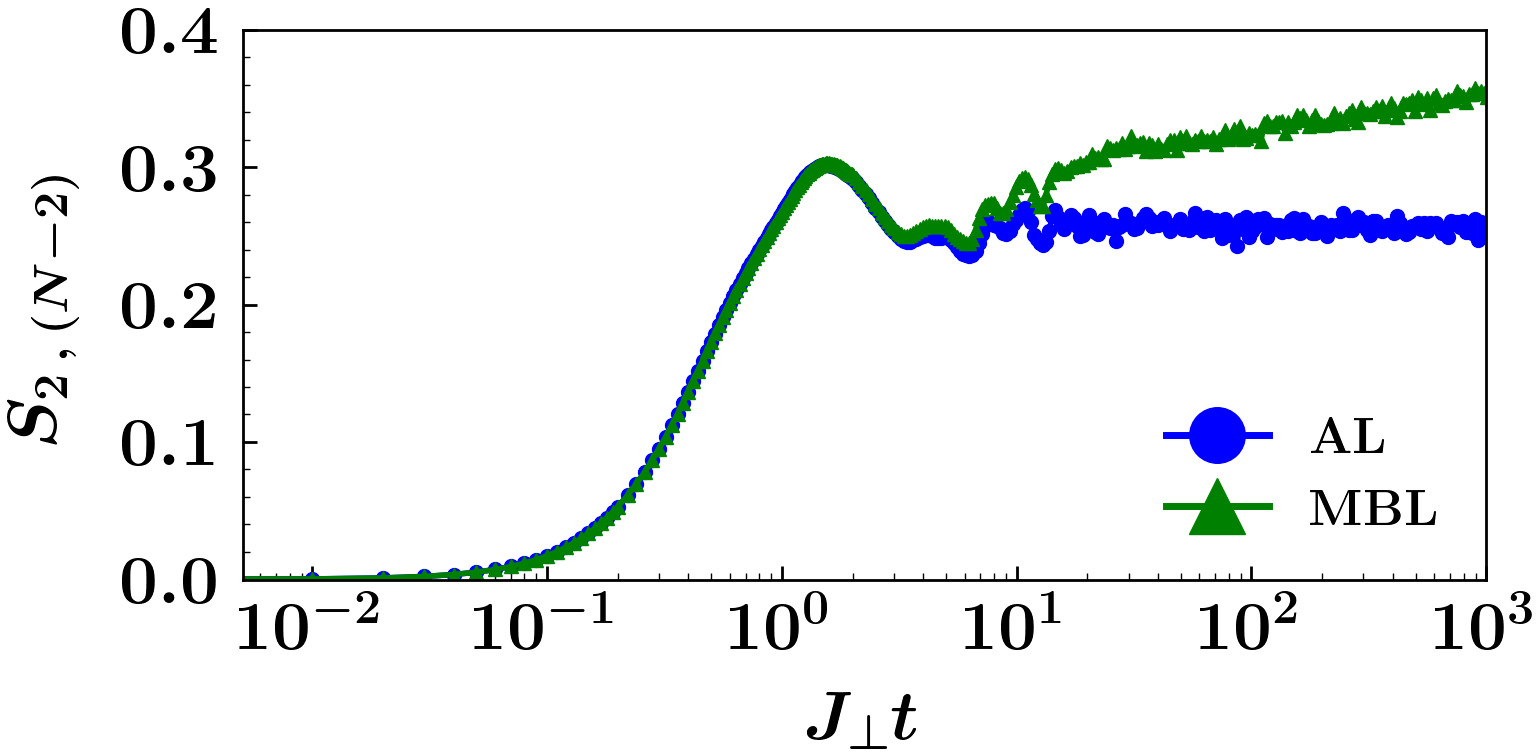}
   \put(-2.5, 47){\scalebox{1.5}{\textbf{(b)}}}  
  \end{overpic}
   \caption{\justifying (a) Quantum spin chain divided into the subsystem $S$ (first two spins), and environment $E$ (remaining spins). The initial Néel state is depicted. (b) Entanglement entropy as a function of time, for the $2,(N-2)$ partition, comparing the MBL phase ($J_z/J_{\perp}=0.2$, green triangle) and AL phase ($J_z/J_{\perp}=0.0$, blue circle). Results are obtained by averaging over $10^3$ disorder realizations, with chain length $N=8$, and disorder strength $W/J_{\perp}=5$.}
  \label{fig:entanglement}
\end{figure}

We preferably choose the first two spins of the chain ($i=1,2$) as subsystem $S$, with the remaining $N-2$ spin to be the environment $E$. The interaction energy between $S$ and $E$ is then related to the bond between the spins at $i=2$ and $i=3$ (see upper panel of Fig.\,\ref{fig:entanglement}). The main aim is to investigate whether localization effects can be inferred by actions on just two spins, representing the minimal subsystem with interacting degrees of freedom \cite{Iemini2016}.
This, for instance, differs from the analytical study of local ergotropy of Ref.\,\cite{SalviaGiovannetti} on the same model, where the subsystem $S$ was identified with the first spin of the chain, under some approximations and without focusing on the different phases of the model.

We perform numerical simulations using MPS techniques to represent the quantum states and operators from ITensor library \cite{fishman2022itensor}. For all the simulations presented in this work, the system is initialized in the Néel state: $\ket{\psi}=\ket{\uparrow, \downarrow,  \uparrow, \downarrow, \dots, \uparrow, \downarrow }$, typically used for MBL dynamics \cite{Abanin, Sierant}. Indeed, Néel state has high mean energy in comparison with the ground state and it is a product state in the spin basis.
Adopting the 2-TDVP \cite{haegeman2011time, haegeman2016unifying}, we compute the time-evolved quantum state. Since the model is disordered, every simulation starts from a different stochastically extracted configuration of disorder. The quantities at each time step $t$ are obtained by averaging over all the realizations of the disorder. 
A closed formula for the local ergotropy is known only for a subsystem made of a single spin-1/2 \cite{SalviaGiovannetti}. Therefore, for our subsystem with two 1/2 spins, we approach the problem numerically. For each realization of disorder and time step, we use a BOA to approximate the optimal local operator for the work extraction task \cite{suppmat}. The approach mimics in spirit the one of Ref.\,\cite{Grazia} and, rigorously speaking, leads to a lower bound for local ergotropy (for simplicity, also referred as local ergotropy in the numerical results).  

{\it Results.}---The entanglement entropy for an isolated system partitioned into two interacting subsystems $S$ and $E$ is defined as the Von-Neumann (VN) entropy of one of these two:
\begin{equation}
\begin{split}
       S_{ent}(SE) &= -\Tr_S\big\{\hat{\rho}_S(t) \ln{\hat{\rho}_S(t)}\big\}=S_{VN}(t).
\end{split}
\end{equation}
being $\hat{\rho}_{S}(t)=\Tr_{E}\hspace{1mm}\big\{\hat{\rho}(t)\big\}$ the subsystem density operator.
It is well known in the literature \cite{Abanin,serbyn2013universal,vedi_parametri} that, for a quantum spin chain partitioned into two equal halves ($N/2 , N/2$), the time evolution of entanglement entropy is a marker for MBL and AL phases. Localized systems show a time behaviour logarithmically dependent on the strength of the interactions: $S_{ent}(t) \sim  \ln( J_z t/\hbar)$, \cite{Abanin,eisert-arealaw,universal-slow-growth,zeng2023logarithmic,toniolo2024dynamics}. Conversely, for ergodic systems, a ballistic growth over time is observed: $S_{ent}(t) \sim  v\hspace{0.5mm}t$. Further details on time-dependent entanglement entropy for this partition are provided in the End Matter, where the ERG case is analyzed, finding perfect agreement with established results.

The following numerical results analyze the case in which the partition is different. As depicted in the upper panel of Fig.\,\ref{fig:entanglement}, our subsystem now consists of $2$ spins, while the remaining $N-2$ sites form the environment. In the lower panel of Fig.\,\ref{fig:entanglement}, we find a qualitative agreement between the entropy dynamics of this two-spin partition and the standard $N/2,N/2$ case, which is included for comparison in Appendix \ref{app:ent_entr} of End Matter.
At initial time, the Néel state has zero entanglement entropy. Then, there is an initial quick rise of the entanglement, until $J_{\perp}t \simeq 1$, corresponding to the expansion of wave packets to a size comparable to the localization length. After this expansion, in the AL phase ($J_z=0$), the entanglement entropy saturates to a constant, but, in the MBL phase ($J_z \ne 0$), it increases logarithmically for any non-zero interaction strength. As discussed in Appendix \ref{app:length} of End Matter, these results are robust upon increasing the chain size $N$.

%
We show now how local ergotropy can be used as a direct probe for discerning ergodic and localized phases. 
One can consider a unitary quench on the subsystem $S$ while this is still interacting with the environment. 
Upon optimization, this approach leads to the recently investigated \cite{Grazia, perciavalle2024extractable} {\it local ergotropy} introduced in Ref.\,\cite{SalviaGiovannetti}, that reads as,
\begin{equation}
\begin{split}
    &\mathcal{E}_S\hspace{0.5mm}[\hat{\rho}(t), \hat{H}] \coloneqq 
    \\
    &\max\limits_{\hat{U}_{S} \in \hspace{0.5mm}\mathcal{U}(d_S)}  \Tr{\hat{H}\bigg(\hat{\rho}(t)-(\hat{U}_S \otimes \hat{I}_E)\hspace{0.5mm}\hat{\rho} (t)\hspace{0.5mm}(\hat{U}_S^{\dag} \otimes \hat{I}_E)\bigg)},
\end{split}
\label{local_erg}
\end{equation}
and is a time-dependent quantity (see also \cite{castellano2024ELE,castellano2025PE} for more recent extended versions). 

\begin{figure}[H]
  \includegraphics[width=1\columnwidth]{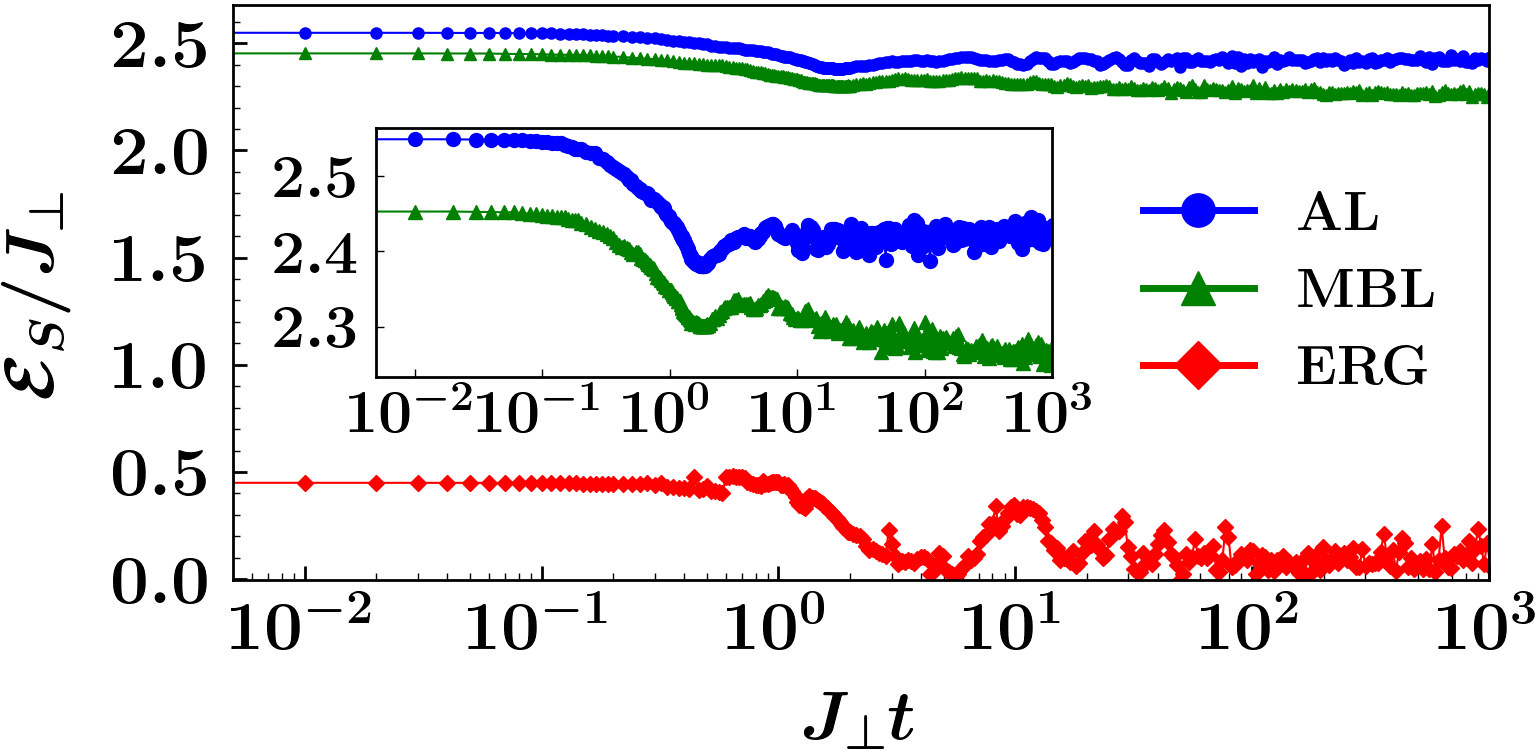}
\caption{\justifying Local ergotropy as a function of time in the three phases: AL (blue circle) for $J_z/J_{\perp}=0.0$ and $W/J_{\perp}=5$; MBL (green triangle) for $J_z/J_{\perp}=0.2$ and $W/J_{\perp}=5$; and ERG (red diamond) for $J_z=0.2$ and $W=0$. The inset provides a zoomed-in view of the local ergotropy as a function of time for the two localized phases. Results are obtained by averaging over $10^3$ disorder realizations, with chain length $N=8$.}
  \label{fig:local-ergo}
\end{figure}

In Fig.\,\ref{fig:local-ergo}, we compare the local ergotropy as a function of time in the three different phases: AL, MBL, and ERG.
Since the system is initially prepared in the high-energy Néel state, we can interpret the subsystem as a charged battery. The post-work extraction state depends on the specific procedure responsible for work extraction. We have mostly simulated the system in the regime $W/J_{\perp}\gg 1$, which ensures that the system is localized and the disorder energy scale dominates the dynamics.  

Fig.\,\ref{fig:local-ergo} shows that in the ERG case ($W=0$, no disorder), after a time transient analogous to that observed in the entanglement entropy ($J_{\perp}t\simeq 1$), the local ergotropy rapidly decreases to zero. This decay is characterized by small oscillations due to finite-size effects (more on this in Appendix \ref{app:length} of End Matter \cite{suppmat}). Consequently, ergodic batteries undergo a rapid discharge, rendering them incapable of sustaining work extraction over long durations.
In contrast, within localized phases, local ergotropy remains much higher. The order of magnitude of the local ergotropy is about $W/2$ in the regime $W/J_{\perp}\gg 1$ \cite{suppmat}. It can be concluded that, within the observed time scales, localized batteries facilitate the energy storage for a much longer duration and the extraction of a greater amount of work in comparison with ergodic batteries.

Focusing now specifically on the difference between the two localized phases, it can be observed that after the initial transient, the spins entangle only weakly with each other, as characteristic of localization. From this point on, a clear distinction emerges between the AL and MBL phases: in the AL phase, local ergotropy stabilizes at a constant value, just like the entanglement entropy. On the other hand, in the MBL case, local ergotropy shows a logarithmic decrease over time, in a manner analogous yet opposite to the entanglement entropy, which instead logarithmically increases. It is thus evident that the time behaviour of local ergotropy can be used as witness for localization phenomena (see also End Matter and \cite{suppmat}). 
Indeed,
\begin{equation}
    \begin{split}
        \mathcal{E}_S=(e^{c}_S+ e^{c}_{int}+e^{c}_E)-(e^{f}_S+e^{f}_{int}+e^{f}_E),
    \end{split}
\end{equation}
where $e^{c}_{S}$ ($e^{f}_{S}$) represents the energy of the subsystem $S$, $e^{c}_{E}$ ($e^{f}_{E}$) is the energy of the environment $E$, and $e^{c}_{Eint}$ ($e^{f}_{int}$) the interaction energy between $S$ and $E$, for the global charged (post-work extraction) state. While the difference $e^{c}_E-e^{f}_E$ can be analytically checked that is zero \cite{SalviaGiovannetti}, in localized phases also the difference $e^{c}_{int}-e^{f}_{int}$ is expected to be small.
Therefore, in localized phases, the local ergotropy is controlled by the difference $e^{c}_{S}-e^{f}_{S}$ derived at each time step from the global system evolution. Nevertheless, local ergotropy turns out to be a dynamical witness of the system phases. 
Thus it can serve as a marker of localization dynamics, directly leveraging local control rather than local projective measurements or entropies.

The second investigated procedure for work extraction involves disconnecting the subsystem from the environment before extracting work from the subsystem $S$. This procedure has the energy cost
\begin{equation}
    \Delta_{SO}(t)= - \Tr_{SE}\hspace{1mm}\big\{\hat{V}_{SE} \hspace{1mm} \hat{\rho}(t)\big\},
    \label{Delta_SO}
\end{equation}
where $\Tr$ denotes the trace operation.
The subsystem ergotropy is the maximum amount of work extractable via unitary operations $\hat{U}_{S}$ acting only on the subsystem $S$, which is now isolated:
\begin{equation}
    \mathcal{E}_{SS}\hspace{0.2mm}[\hat{\rho}_{S}(t), \hat{H}_{S}] \coloneqq \max\limits_{\hat{U}_{S} \in \hspace{0.2mm}\mathcal{U}(d_S)}  \Tr_S 
    {\hat{H}_{S}\bigg(\hat{\rho}_{S}(t)-\hat{U}_{S}\hspace{0.5mm}\hat{\rho}_{S}(t) \hspace{0.5mm}\hat{U}_{S}^{\dag}\bigg)}.
\end{equation}
This is also a time dependent quantity.
Finally, the total amount of work extractable from $S$, including the previous disconnection, is defined as {\it switch-off ergotropy} \cite{SalviaGiovannetti},
\begin{equation}
    \mathcal{E}_{SO} \hspace{0.5mm}[\hat{\rho}(t), \hat{H}]   \coloneqq \mathcal{E}_{SS}\hspace{0.5mm}[\hat{\rho}_{S}(t), \hat{H}_{S}]-\Delta_{SO}(t).
\end{equation}
One important consideration is that turning off the interactions implicitly requires control over $E$ as well. 


We compare the two aforementioned protocols for local and switch-off ergotropy. 
\begin{figure}[htbp!]
  \includegraphics[width=1\columnwidth]{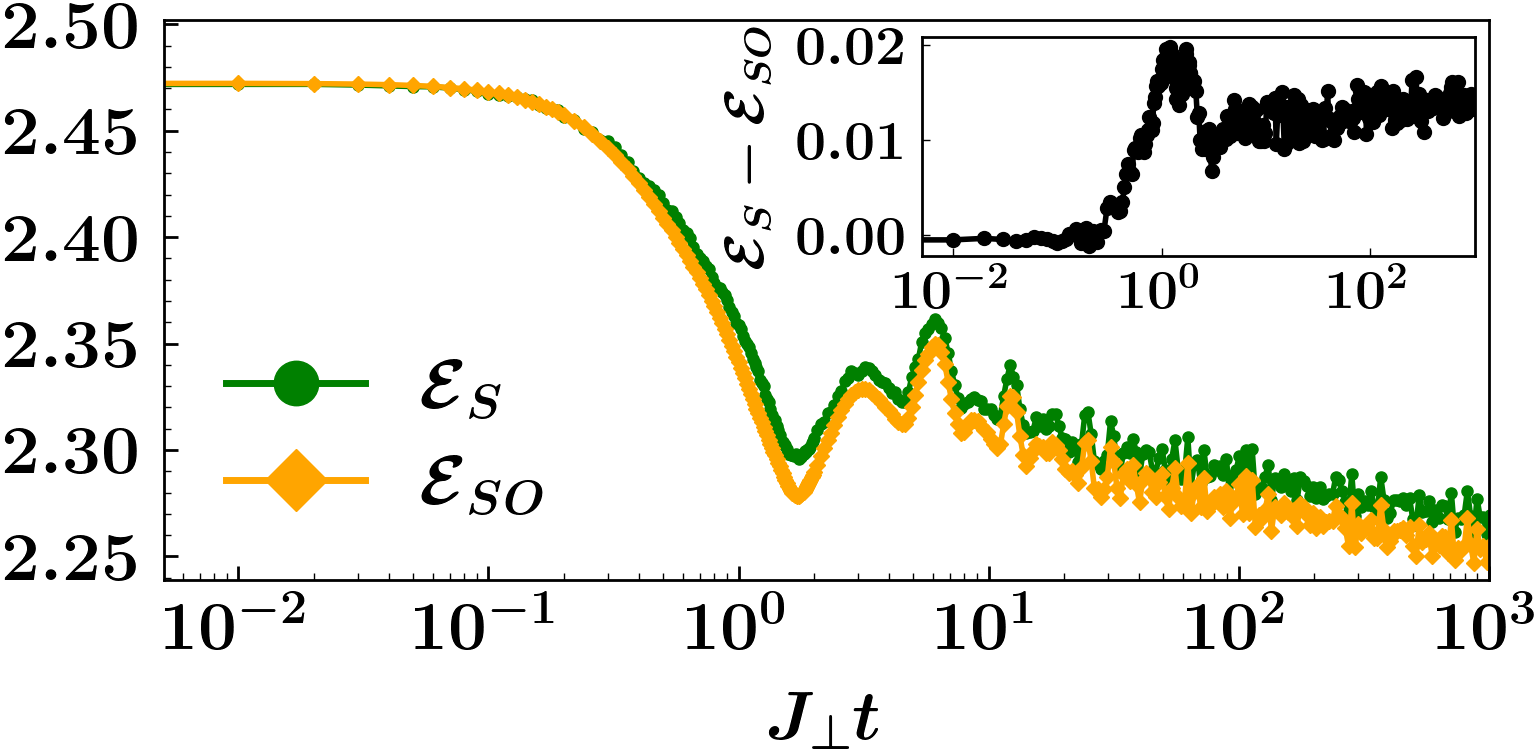}
  \caption{\justifying Local ergotropy $\mathcal{E}_{S}$ (green circle) and switch-off ergotropy $\mathcal{E}_{SO}$ (orange triangle), both in units of $J_{\perp}$, as functions of time for the system in MBL phase. 
  The inset provides a view over the difference between local ergotropy and the switch-off ergotropy $\mathcal{E}_{S}-\mathcal{E}_{SO}$. 
  Results are obtained by averaging over $2.6\cdot 10^3$ disorder realizations, with chain length $N=12$, disorder strength $W/J_{\perp}=5$, and $J_z/J_{\perp}=0.2$.}
   \label{fig:local-vs-SO-ergo}
\end{figure}
The plot in Fig.\,\ref{fig:local-vs-SO-ergo} for the switch-off ergotropy can be related to the results previously found in \cite{Polini}. Moreover, it reveals that the switch-off ergotropy is smaller than the local ergotropy, albeit slightly. This is because the energy term $\Delta_{SO}$ in Eq.\,\eqref{Delta_SO} is small in localized phases. Since the longitudinal interactions are small compared to the other energy scales ($J_z \ll W,J_{\perp}$), the difference between the local and the switch-off ergotropy remains small in localized phases. We also perform work extraction without BOA, observing that it is equivalent to the switch-off protocol, allowing a gain of $1\%$ on the local ergotropy value \cite{suppmat}. 


\begin{figure}[!htbp]
  \includegraphics[width=1\columnwidth]{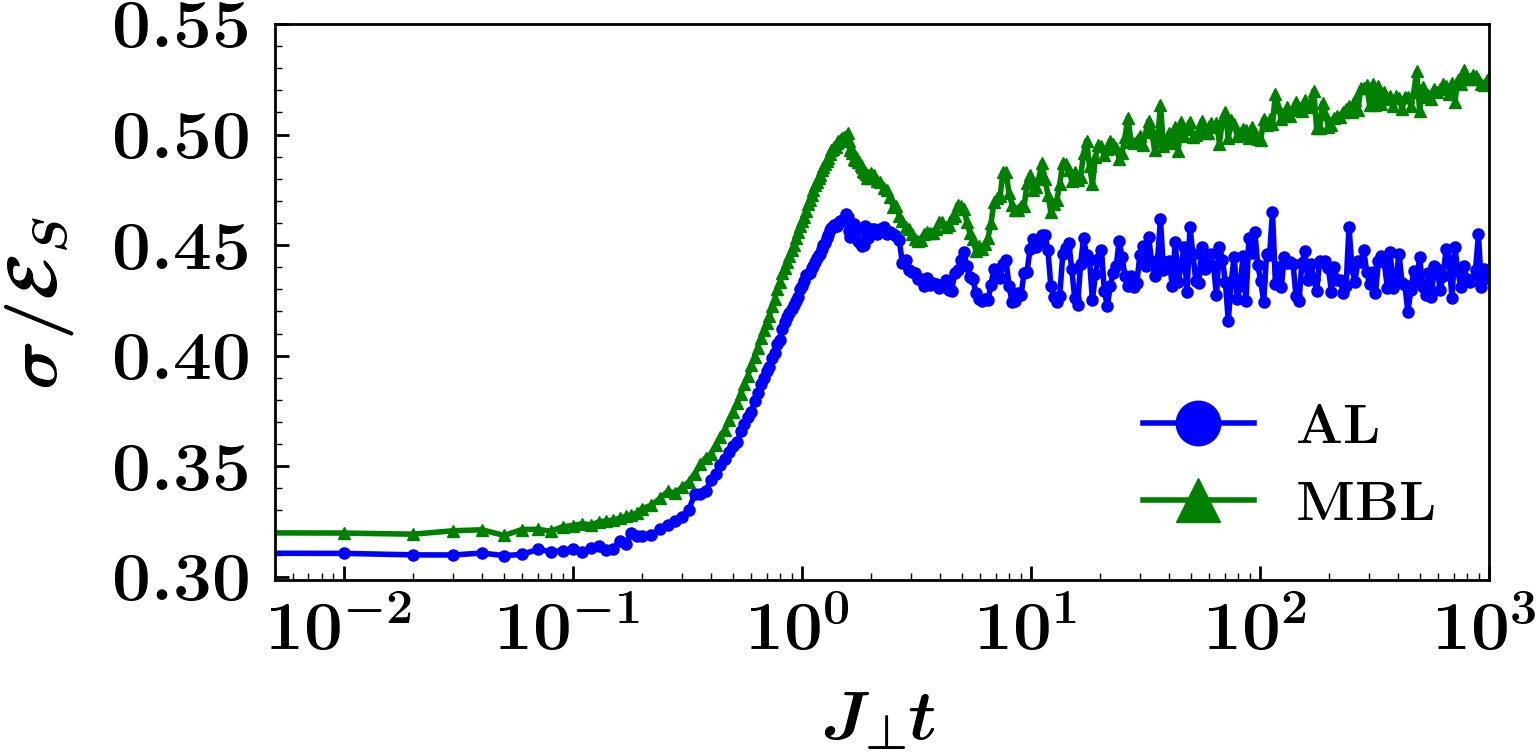}
  \caption{\justifying Relative quantum fluctuations of the local ergotropy as a function of time in the MBL ($J_z/J_{\perp}=0.2$, green triangle) and AL phase ($J_z/J_{\perp}=0.0$, blue circle). Results are obtained by averaging over $10^3$ disorder realizations for the AL case and approximately $2.6 \cdot 10^3$ disorder realizations for the MBL case, both with chain length $N=8$, and disorder strength $W/J_{\perp}=5$.}
\label{fig:fluttuazioni_quantistiche}
\end{figure}

We have finally computed quantum fluctuations of the local ergotropy following Ref.\,\cite{Quantum-correlations-francica}:
\begin{equation}
    \begin{split}
        \sigma^2(t) = \Tr[(\hat{H'}-\hat{H})^2\hat{\rho}(t)]-
        \left\{\Tr[(\hat{H'}-\hat{H})\hat{\rho}(t)]\right\}^2,
    \end{split}
\end{equation}
in which $\hat{H}'=\hat{U}_S^{\dag}\hat{H}\hat{U}_S$. Numerical results in Fig.\,\ref{fig:fluttuazioni_quantistiche} show that the fluctuations behaviour over time also represents a marker for MBL and AL phases. In analogy with the behaviour of the entanglement entropy, after a transient of ballistic growth, the fluctuations saturate to a constant value in the AL case, while they grow logarithmically in the MBL case. For both phases, the short-time value is due to the quantum fluctuations of the initial state and the disorder strength. In the Appendix \ref{app:quantum_fluc} of End Matter, we show that increasing the disorder strength $W$ enhances the value of local ergotropy while, remarkably, reducing relative quantum fluctuations. Disorder-induced fluctuations turn out to be smaller than quantum fluctuations \cite{suppmat}. We find that disorder fluctuations are larger in the AL phase than in MBL phase, consistent with previous results in the literature \cite{Polini}. Moreover, while disorder fluctuations increase with $W$, the relative disorder fluctuations remain nearly constant, unlike quantum fluctuations.

{\it Conclusions.}---We have demonstrated that local ergotropy and its quantum fluctuations can serve as thermodynamic markers of localization phenomena. Specifically, in the ERG phase, the extractable work rapidly drops to zero, in the AL phase, it reaches a stationary constant value, and, in MBL case, it decreases logarithmically over time. A summary of the different dynamical signatures of localization phenomena is reported in Table \ref{tabella}. 
Our findings open the way to characterization of different localized phases in many-body systems via direct local work extraction in the presence of Hamiltonian coupling. 
The analysis presented in this work can be extended to MBL systems in the presence of an electric field without disorder \cite{taylor2020experimental} and generalized to other models, such as the Sachdev–Ye–Kitaev one \cite{campaioli2024colloquium,rossini2020} and open ones \cite{de2023signatures,di2024environment}.

{\it Acknowledgements.---}D.F. and G.D.F. acknowledge financial support from PNRR MUR Project No. PE0000023-NQSTI. C.A.P. acknowledges funding from IQARO (Spin-orbitronic Quantum Bits in Reconfigurable 2DOxides)
project of the European Union’s Horizon Europe research and innovation programme under grant agreement n. 101115190. G.D.B., G.D.F. and C.A.P. acknowledge funding from the PRIN 2022 project 2022FLSPAJ ``Taming Noisy Quantum Dynamics'' (TANQU). C.A.P. acknowledges funding from the PRIN 2022 PNRR project P2022SB73K ``Superconductivity in KTaO3 Oxide-2DEG NAnodevices for Topological quantum Applications'' (SONATA) financed by the European Union - Next Generation EU. The authors acknowledge stimulating discussions with A. Russomanno, P. Lucignano, and M. Sassetti.


\begin{table}[H]
\centering
\vspace{0.5cm}
\begin{NiceTabular}{l | c | c | c }[colortbl-like]
    \toprule
     &
    \textcolor{rosso}{\textbf{ERG}}&
    \textcolor{blu}{\textbf{AL}}&
    \textcolor{verde}{\textbf{MBL}}\\
    \midrule
    $S_{({N/2,N/2})}$ & \cellcolor{giallo}\textbf{Ballistic} & \cellcolor{giallo}\textbf{Constant} & \cellcolor{giallo}\textbf{Logarithmic} \\
    & \cellcolor{giallo}\textbf{increase} & \cellcolor{giallo} & \cellcolor{giallo}\textbf{increase} \\
    \midrule
    $S_{2,
    ({N-2})}$ & \cellcolor{rosso}\textcolor{white}{\textbf{Ballistic}}
 & \cellcolor{blu}\textcolor{white}{\textbf{Constant}} & \cellcolor{verde}\textcolor{white}{\textbf{Logarithmic}} \\
    & \cellcolor{rosso}\textcolor{white}{\textbf{increase}} & \cellcolor{blu} & \cellcolor{verde}\textcolor{white}{\textbf{increase}} \\
    \midrule
    $\mathcal{E}_{S}$ & \cellcolor{rosso}\textcolor{white}{\textbf{Ballistic}} & \cellcolor{blu}\textcolor{white}{\textbf{Constant}} & \cellcolor{verde}\textcolor{white}{\textbf{Logarithmic}} \\
    & \cellcolor{rosso}\textcolor{white}{\textbf{decrease}} & \cellcolor{blu} & \cellcolor{verde}\textbf{\textcolor{white}{decrease}} \\
    \midrule
    $\sigma$ & \cellcolor{rosso}\textcolor{white}{\textbf{Ballistic}} & \cellcolor{blu}\textcolor{white}{\textbf{Constant}} & \cellcolor{verde}\textcolor{white}{\textbf{Logarithmic}} \\
    & \cellcolor{rosso}\textcolor{white}{\textbf{increase}} & \cellcolor{blu} & \cellcolor{verde}\textcolor{white}{\textbf{increase}} \\
    \bottomrule
\end{NiceTabular}
\caption{\label{tab:fid}\justifying Dynamical signatures of localization phenomena (yellow cells for established previous results).}
\label{tabella}
\end{table}

\bibliographystyle{ieeetr}
\bibliography{Bibliografia.bib}

\begin{thebibliography}{10}

\bibitem{Nandki}
R.~Nandkishore and D.~A. Huse, ``Many-body localization and thermalization in quantum statistical mechanics,'' {\em Annu. Rev. Condens. Matter Phys.}, vol.~6, pp.~15--38, 2015.

\bibitem{Abanin}
D.~A. Abanin, E.~Altman, I.~Bloch, and M.~Serbyn, ``Many-body localization, thermalization, and entanglement,'' {\em Rev. Mod. Phys.}, vol.~91, p.~021001, 2019.

\bibitem{Alet}
F.~Alet and N.~Laflorencie, ``Many-body localization: An introduction and selected topics,'' {\em C. R. Phys.}, vol.~19, pp.~498--525, 2018.

\bibitem{Sierant}
P.~Sierant, M.~Lewenstein, A.~Scardicchio, L.~Vidmar, and J.~Zakrzewski, ``Many-body localization in the age of classical computing,'' {\em Rep. Prog. Phys.}, vol.~88, p.~026502, 2025.

\bibitem{serbyn2013universal}
M.~Serbyn, Z.~Papi{\'c}, and D.~A. Abanin, ``Universal slow growth of entanglement in interacting strongly disordered systems,'' {\em Phys. Rev. Lett.}, vol.~110, no.~26, p.~260601, 2013.

\bibitem{vedi_parametri}
J.~H. Bardarson, F.~Pollmann, and J.~E. Moore, ``Unbounded growth of entanglement in models of many-body localization,'' {\em Phys. Rev. Lett.}, vol.~109, no.~1, p.~017202, 2012.

\bibitem{lukin2019probing}
A.~Lukin, M.~Rispoli, R.~Schittko, M.~E. Tai, A.~M. Kaufman, S.~Choi, V.~Khemani, J.~L{\'e}onard, and M.~Greiner, ``Probing entanglement in a many-body--localized system,'' {\em Sci.}, vol.~364, no.~6437, pp.~256--260, 2019.

\bibitem{schreiber2015observation}
M.~Schreiber, S.~S. Hodgman, P.~Bordia, H.~P. L{\"u}schen, M.~H. Fischer, R.~Vosk, E.~Altman, U.~Schneider, and I.~Bloch, ``Observation of many-body localization of interacting fermions in a quasirandom optical lattice,'' {\em Sci.}, vol.~349, no.~6250, pp.~842--845, 2015.

\bibitem{smith2016many}
J.~Smith, A.~Lee, P.~Richerme, B.~Neyenhuis, P.~W. Hess, P.~Hauke, M.~Heyl, D.~A. Huse, and C.~Monroe, ``Many-body localization in a quantum simulator with programmable random disorder,'' {\em Nat. Phys.}, vol.~12, no.~10, pp.~907--911, 2016.

\bibitem{Choi}
J.~Choi, S.~Hild, J.~Zeiher, P.~Schauß, A.~Rubio-Abadal, T.~Yefsah, V.~Khemani, D.~A. Huse, I.~Bloch, and C.~Gross, ``Exploring the many-body localization transition in two dimensions,'' {\em Sci.}, vol.~352, p.~1547, 2016.

\bibitem{Kohlert}
T.~Kohlert, S.~Scherg, X.~Li, H.~P. Lüschen, S.~Das~Sarma, I.~Bloch, and M.~A. M, ``Observation of many-body localization in a one-dimensional system with a single-particle mobility edge,'' {\em Phys. Rev. Lett.}, vol.~122, p.~170403, 2019.

\bibitem{gong2021experimental}
M.~Gong, G.~D. de~Moraes~Neto, C.~Zha, Y.~Wu, H.~Rong, Y.~Ye, S.~Li, Q.~Zhu, S.~Wang, Y.~Zhao, {\em et~al.}, ``Experimental characterization of the quantum many-body localization transition,'' {\em Phys. Rev. Res.}, vol.~3, no.~3, p.~033043, 2021.

\bibitem{campaioli2024colloquium}
F.~Campaioli, S.~Gherardini, J.~Q. Quach, M.~Polini, and G.~M. Andolina, ``Quantum batteries,'' {\em Rev. Mod. Phys.}, vol.~96, no.~3, p.~031001, 2024.

\bibitem{le2018spin}
T.~P. Le, J.~Levinsen, K.~Modi, M.~M. Parish, and F.~A. Pollock, ``Spin-chain model of a many-body quantum battery,'' {\em Phys. Rev. A}, vol.~97, no.~2, p.~022106, 2018.

\bibitem{ferraro2018high}
D.~Ferraro, M.~Campisi, G.~M. Andolina, V.~Pellegrini, and M.~Polini, ``High-power collective charging of a solid-state quantum battery,'' {\em Phys. Rev. Lett.}, vol.~120, no.~11, p.~117702, 2018.

\bibitem{andolina2018charger}
G.~M. Andolina, D.~Farina, A.~Mari, V.~Pellegrini, V.~Giovannetti, and M.~Polini, ``Charger-mediated energy transfer in exactly solvable models for quantum batteries,'' {\em Phys. Rev. B}, vol.~98, no.~20, p.~205423, 2018.

\bibitem{farina2019}
D.~Farina, G.~M. Andolina, A.~Mari, M.~Polini, and V.~Giovannetti, ``Charger-mediated energy transfer for quantum batteries: An open-system approach,'' {\em Phys. Rev. B}, vol.~99, no.~035421, 2019.

\bibitem{mitchison2021charging}
M.~T. Mitchison, J.~Goold, and J.~Prior, ``Charging a quantum battery with linear feedback control,'' {\em Quantum}, vol.~5, p.~500, 2021.

\bibitem{dou2022cavity}
F.-Q. Dou, H.~Zhou, and J.-A. Sun, ``Cavity {H}eisenberg-spin-chain quantum battery,'' {\em Phys. Rev. A}, vol.~106, no.~3, p.~032212, 2022.

\bibitem{yao2022optimal}
Y.~Yao and X.~Shao, ``Optimal charging of open spin-chain quantum batteries via homodyne-based feedback control,'' {\em Phys. Rev. E}, vol.~106, no.~1, p.~014138, 2022.

\bibitem{mazzoncini2023optimal}
F.~Mazzoncini, V.~Cavina, G.~M. Andolina, P.~A. Erdman, and V.~Giovannetti, ``Optimal control methods for quantum batteries,'' {\em Phys. Rev. A}, vol.~107, no.~3, p.~032218, 2023.

\bibitem{andolina2019}
G.~M. Andolina, M.~Keck, A.~Mari, M.~Campisi, V.~Giovannetti, and M.~Polini, ``Extractable work, the role of correlations, and asymptotic freedom in quantum batteries,'' {\em Phys. Rev. Lett.}, vol.~122, no.~047702, 2019.

\bibitem{arjmandi2022performance}
M.~B. Arjmandi, A.~Shokri, E.~Faizi, and H.~Mohammadi, ``Performance of quantum batteries with correlated and uncorrelated chargers,'' {\em Phys. Rev. A}, vol.~106, no.~6, p.~062609, 2022.

\bibitem{joshi2022experimental}
J.~Joshi and T.~Mahesh, ``Experimental investigation of a quantum battery using star-topology {NMR} spin systems,'' {\em Phys. Rev. A}, vol.~106, no.~4, p.~042601, 2022.

\bibitem{hu2022optimal}
C.-K. Hu, J.~Qiu, P.~J. Souza, J.~Yuan, Y.~Zhou, L.~Zhang, J.~Chu, X.~Pan, L.~Hu, J.~Li, {\em et~al.}, ``Optimal charging of a superconducting quantum battery,'' {\em Quantum Sci. and Technol.}, vol.~7, no.~4, p.~045018, 2022.

\bibitem{allahverdyan}
A.~E. Allahverdyan, R.~Balian, and T.~M. Nieuwenhuizen, ``Maximal work extraction from finite quantum systems,'' {\em Europhys. Lett.}, vol.~67, no.~4, p.~565, 2004.

\bibitem{alicki2013entanglement}
R.~Alicki and M.~Fannes, ``Entanglement boost for extractable work from ensembles of quantum batteries,'' {\em Phys. Rev. E}, vol.~87, no.~4, p.~042123, 2013.

\bibitem{lenard1978thermodynamical}
A.~Lenard, ``Thermodynamical proof of the {G}ibbs formula for elementary quantum systems,'' {\em J. Stat. Phys.}, vol.~19, pp.~575--586, 1978.

\bibitem{pusz1978passive}
W.~Pusz and S.~L. Woronowicz, ``Passive states and kms states for general quantum systems,'' {\em Commun. Math. Phys.}, vol.~58, pp.~273--290, 1978.

\bibitem{Polini}
D.~Rossini, G.~M. Andolina, and M.~Polini, ``Many-body localized quantum batteries,'' {\em Phys. Rev. B}, vol.~100, no.~11, p.~115142, 2019.

\bibitem{SalviaGiovannetti}
R.~Salvia, G.~De~Palma, and V.~Giovannetti, ``Optimal local work extraction from bipartite quantum systems in the presence of {H}amiltonian couplings,'' {\em Phys. Rev. A}, vol.~107, no.~1, p.~012405, 2023.

\bibitem{murmann2015antiferromagnetic}
S.~Murmann, F.~Deuretzbacher, G.~Z{\"u}rn, J.~Bjerlin, S.~M. Reimann, L.~Santos, T.~Lompe, and S.~Jochim, ``Antiferromagnetic {H}eisenberg spin chain of a few cold atoms in a one-dimensional trap,'' {\em Phys. Rev. Lett.}, vol.~115, no.~21, p.~215301, 2015.

\bibitem{perez2006matrix}
D.~Perez-Garcia, F.~Verstraete, M.~M. Wolf, and J.~I. Cirac, ``Matrix product state representations,'' {\em arXiv preprint quant-ph/0608197}, 2006.

\bibitem{white1992density}
S.~R. White, ``Density matrix formulation for quantum renormalization groups,'' {\em Phys. Rev. Lett.}, vol.~69, no.~19, p.~2863, 1992.

\bibitem{schollwock2011density}
U.~Schollw{\"o}ck, ``The density-matrix renormalization group in the age of matrix product states,'' {\em Ann. Phys.}, vol.~326, no.~1, pp.~96--192, 2011.

\bibitem{haegeman2011time}
J.~Haegeman, J.~I. Cirac, T.~J. Osborne, I.~Pi{\v{z}}orn, H.~Verschelde, and F.~Verstraete, ``Time-dependent variational principle for quantum lattices,'' {\em Phys. Rev. Lett.}, vol.~107, no.~7, p.~070601, 2011.

\bibitem{haegeman2016unifying}
J.~Haegeman, C.~Lubich, I.~Oseledets, B.~Vandereycken, and F.~Verstraete, ``Unifying time evolution and optimization with matrix product states,'' {\em Phys. Rev. B}, vol.~94, no.~16, p.~165116, 2016.

\bibitem{frazier2018tutorial}
P.~I. Frazier, ``A tutorial on {B}ayesian optimization,'' {\em arXiv preprint arXiv:1807.02811}, 2018.

\bibitem{grazi2024prldimer}
R.~Grazi, D.~Sacco~Shaikh, M.~Sassetti, N.~Traverso~Ziani, and D.~Ferraro, ``Controlling energy storage crossing quantum phase transitions in an integrable spin quantum battery,'' {\em Phys. Rev. Lett.}, vol.~133, p.~197001, Nov 2024.

\bibitem{suppmat}
{\rm See Supplemental Material for more details.}

\bibitem{Iemini2016}
F.~Iemini, A.~Russomanno, D.~Rossini, A.~Scardicchio, and R.~Fazio, ``Signatures of many-body localization in the dynamics of two-site entanglement,'' {\em Phys. Rev. B}, vol.~94, no.~214206, 2016.

\bibitem{fishman2022itensor}
M.~Fishman, S.~R. White, and E.~M. Stoudenmire, ``{The ITensor Software Library for Tensor Network Calculations},'' {\em SciPost Phys. Codebases}, p.~4, 2022.

\bibitem{Grazia}
G.~Di~Bello, D.~Farina, D.~Jansen, C.~A. Perroni, V.~Cataudella, and G.~De~Filippis, ``Local ergotropy and its fluctuations across a dissipative quantum phase transition,'' {\em Quantum Sci. and Technol.}, vol.~10, no.~1, p.~015049, 2024.

\bibitem{eisert-arealaw}
J.~Eisert, M.~Cramer, and M.~B. Plenio, ``Area laws for the entanglement entropy-a review,'' {\em arXiv preprint arXiv:0808.3773}, 2008.

\bibitem{universal-slow-growth}
M.~Serbyn, Z.~Papi{\'c}, and D.~A. Abanin, ``Universal slow growth of entanglement in interacting strongly disordered systems,'' {\em Phys. Rev. Lett.}, vol.~110, no.~26, p.~260601, 2013.

\bibitem{zeng2023logarithmic}
Y.~Zeng, A.~Hamma, Y.-R. Zhang, Q.~Liu, R.~Li, H.~Fan, and W.-M. Liu, ``Logarithmic light cone, slow entanglement growth, and quantum memory,'' {\em arXiv preprint arXiv:2305.08334}, 2023.

\bibitem{toniolo2024dynamics}
D.~Toniolo and S.~Bose, ``Dynamics of many-body localized systems: logarithmic lightcones and $\log$, t-law of $\alpha$-{R}{\'e}nyi entropies,'' {\em arXiv preprint arXiv:2408.02016}, 2024.

\bibitem{perciavalle2024extractable}
F.~Perciavalle, D.~Rossini, J.~Polo, and L.~Amico, ``Extractable energy from quantum superposition of current states,'' {\em arXiv preprint arXiv:2410.13934}, 2024.

\bibitem{castellano2024ELE}
R.~Castellano, D.~Farina, V.~Giovannetti, and A.~Acin, ``Extended local ergotropy,'' {\em Phys. Rev. Lett.}, vol.~133, p.~150402, Oct 2024.

\bibitem{castellano2025PE}
R.~Castellano, R.~Nery, K.~Simonov, and D.~Farina, ``Parallel ergotropy: Maximum work extraction via parallel local unitary operations,'' {\em Phys. Rev. A}, vol.~111, p.~012212, Jan 2025.

\bibitem{Quantum-correlations-francica}
G.~Francica, ``Quantum correlations and ergotropy,'' {\em Phys. Rev. E}, vol.~105, no.~5, p.~L052101, 2022.

\bibitem{taylor2020experimental}
S.~R. Taylor, M.~Schulz, F.~Pollmann, and R.~Moessner, ``Experimental probes of {S}tark many-body localization,'' {\em Phys. Rev. B}, vol.~102, no.~5, p.~054206, 2020.

\bibitem{rossini2020}
D.~Rossini, G.~M. Andolina, D.~Rosa, M.~Carrega, and M.~Polini, ``Quantum advantage in the charging process of {S}achdev-{Y}e-{K}itaev batteries,'' {\em Phys. Rev. Lett.}, vol.~125, no.~236402, 2020.

\bibitem{de2023signatures}
G.~De~Filippis, A.~De~Candia, G.~Di~Bello, C.~A. Perroni, L.~Cangemi, A.~Nocera, M.~Sassetti, R.~Fazio, and V.~Cataudella, ``Signatures of dissipation driven quantum phase transition in {R}abi model,'' {\em Phys. Rev. Lett.}, vol.~130, no.~21, p.~210404, 2023.

\bibitem{di2024environment}
G.~Di~Bello, A.~Ponticelli, F.~Pavan, V.~Cataudella, G.~De~Filippis, A.~de~Candia, and C.~A. Perroni, ``Environment induced dynamical quantum phase transitions in two-qubit {R}abi model,'' {\em Commun. Phys.}, vol.~7, no.~1, p.~364, 2024.

\end{thebibliography}


\begin{thebibliography}{11}%
\makeatletter
\providecommand \@ifxundefined [1]{%
 \@ifx{#1\undefined}
}%
\providecommand \@ifnum [1]{%
 \ifnum #1\expandafter \@firstoftwo
 \else \expandafter \@secondoftwo
 \fi
}%
\providecommand \@ifx [1]{%
 \ifx #1\expandafter \@firstoftwo
 \else \expandafter \@secondoftwo
 \fi
}%
\providecommand \natexlab [1]{#1}%
\providecommand \enquote  [1]{``#1''}%
\providecommand \bibnamefont  [1]{#1}%
\providecommand \bibfnamefont [1]{#1}%
\providecommand \citenamefont [1]{#1}%
\providecommand \href@noop [0]{\@secondoftwo}%
\providecommand \href [0]{\begingroup \@sanitize@url \@href}%
\providecommand \@href[1]{\@@startlink{#1}\@@href}%
\providecommand \@@href[1]{\endgroup#1\@@endlink}%
\providecommand \@sanitize@url [0]{\catcode `\\12\catcode `\$12\catcode `\&12\catcode `\#12\catcode `\^12\catcode `\_12\catcode `\%12\relax}%
\providecommand \@@startlink[1]{}%
\providecommand \@@endlink[0]{}%
\providecommand \url  [0]{\begingroup\@sanitize@url \@url }%
\providecommand \@url [1]{\endgroup\@href {#1}{\urlprefix }}%
\providecommand \urlprefix  [0]{URL }%
\providecommand \Eprint [0]{\href }%
\providecommand \doibase [0]{https://doi.org/}%
\providecommand \selectlanguage [0]{\@gobble}%
\providecommand \bibinfo  [0]{\@secondoftwo}%
\providecommand \bibfield  [0]{\@secondoftwo}%
\providecommand \translation [1]{[#1]}%
\providecommand \BibitemOpen [0]{}%
\providecommand \bibitemStop [0]{}%
\providecommand \bibitemNoStop [0]{.\EOS\space}%
\providecommand \EOS [0]{\spacefactor3000\relax}%
\providecommand \BibitemShut  [1]{\csname bibitem#1\endcsname}%
\let\auto@bib@innerbib\@empty
\bibitem [{\citenamefont {Rossini}\ \emph {et~al.}(2019)\citenamefont {Rossini}, \citenamefont {Andolina},\ and\ \citenamefont {Polini}}]{Polini}%
  \BibitemOpen
  \bibfield  {author} {\bibinfo {author} {\bibfnamefont {D.}~\bibnamefont {Rossini}}, \bibinfo {author} {\bibfnamefont {G.~M.}\ \bibnamefont {Andolina}},\ and\ \bibinfo {author} {\bibfnamefont {M.}~\bibnamefont {Polini}},\ }\bibfield  {title} {\bibinfo {title} {Many-body localized quantum batteries},\ }\href@noop {} {\bibfield  {journal} {\bibinfo  {journal} {Phys. Rev. B}\ }\textbf {\bibinfo {volume} {100}},\ \bibinfo {pages} {115142} (\bibinfo {year} {2019})}\BibitemShut {NoStop}%
\bibitem [{\citenamefont {Campaioli}\ \emph {et~al.}(2024)\citenamefont {Campaioli}, \citenamefont {Gherardini}, \citenamefont {Quach}, \citenamefont {Polini},\ and\ \citenamefont {Andolina}}]{campaioli2024colloquium}%
  \BibitemOpen
  \bibfield  {author} {\bibinfo {author} {\bibfnamefont {F.}~\bibnamefont {Campaioli}}, \bibinfo {author} {\bibfnamefont {S.}~\bibnamefont {Gherardini}}, \bibinfo {author} {\bibfnamefont {J.~Q.}\ \bibnamefont {Quach}}, \bibinfo {author} {\bibfnamefont {M.}~\bibnamefont {Polini}},\ and\ \bibinfo {author} {\bibfnamefont {G.~M.}\ \bibnamefont {Andolina}},\ }\bibfield  {title} {\bibinfo {title} {Quantum batteries},\ }\href@noop {} {\bibfield  {journal} {\bibinfo  {journal} {Rev. Mod. Phys.}\ }\textbf {\bibinfo {volume} {96}},\ \bibinfo {pages} {031001} (\bibinfo {year} {2024})}\BibitemShut {NoStop}%
\bibitem [{\citenamefont {Perez-Garcia}\ \emph {et~al.}(2006)\citenamefont {Perez-Garcia}, \citenamefont {Verstraete}, \citenamefont {Wolf},\ and\ \citenamefont {Cirac}}]{perez2006matrix}%
  \BibitemOpen
  \bibfield  {author} {\bibinfo {author} {\bibfnamefont {D.}~\bibnamefont {Perez-Garcia}}, \bibinfo {author} {\bibfnamefont {F.}~\bibnamefont {Verstraete}}, \bibinfo {author} {\bibfnamefont {M.~M.}\ \bibnamefont {Wolf}},\ and\ \bibinfo {author} {\bibfnamefont {J.~I.}\ \bibnamefont {Cirac}},\ }\bibfield  {title} {\bibinfo {title} {Matrix product state representations},\ }\href@noop {} {\bibfield  {journal} {\bibinfo  {journal} {arXiv preprint quant-ph/0608197}\ } (\bibinfo {year} {2006})}\BibitemShut {NoStop}%
\bibitem [{\citenamefont {White}(1992)}]{white1992density}%
  \BibitemOpen
  \bibfield  {author} {\bibinfo {author} {\bibfnamefont {S.~R.}\ \bibnamefont {White}},\ }\bibfield  {title} {\bibinfo {title} {Density matrix formulation for quantum renormalization groups},\ }\href@noop {} {\bibfield  {journal} {\bibinfo  {journal} {Phys. Rev. Lett.}\ }\textbf {\bibinfo {volume} {69}},\ \bibinfo {pages} {2863} (\bibinfo {year} {1992})}\BibitemShut {NoStop}%
\bibitem [{\citenamefont {Schollw{\"o}ck}(2011)}]{schollwock2011density}%
  \BibitemOpen
  \bibfield  {author} {\bibinfo {author} {\bibfnamefont {U.}~\bibnamefont {Schollw{\"o}ck}},\ }\bibfield  {title} {\bibinfo {title} {The density-matrix renormalization group in the age of matrix product states},\ }\href@noop {} {\bibfield  {journal} {\bibinfo  {journal} {Ann. Phys.}\ }\textbf {\bibinfo {volume} {326}},\ \bibinfo {pages} {96} (\bibinfo {year} {2011})}\BibitemShut {NoStop}%
\bibitem [{\citenamefont {Haegeman}\ \emph {et~al.}(2011)\citenamefont {Haegeman}, \citenamefont {Cirac}, \citenamefont {Osborne}, \citenamefont {Pi{\v{z}}orn}, \citenamefont {Verschelde},\ and\ \citenamefont {Verstraete}}]{haegeman2011time}%
  \BibitemOpen
  \bibfield  {author} {\bibinfo {author} {\bibfnamefont {J.}~\bibnamefont {Haegeman}}, \bibinfo {author} {\bibfnamefont {J.~I.}\ \bibnamefont {Cirac}}, \bibinfo {author} {\bibfnamefont {T.~J.}\ \bibnamefont {Osborne}}, \bibinfo {author} {\bibfnamefont {I.}~\bibnamefont {Pi{\v{z}}orn}}, \bibinfo {author} {\bibfnamefont {H.}~\bibnamefont {Verschelde}},\ and\ \bibinfo {author} {\bibfnamefont {F.}~\bibnamefont {Verstraete}},\ }\bibfield  {title} {\bibinfo {title} {Time-dependent variational principle for quantum lattices},\ }\href@noop {} {\bibfield  {journal} {\bibinfo  {journal} {Phys. Rev. Lett.}\ }\textbf {\bibinfo {volume} {107}},\ \bibinfo {pages} {070601} (\bibinfo {year} {2011})}\BibitemShut {NoStop}%
\bibitem [{\citenamefont {Haegeman}\ \emph {et~al.}(2016)\citenamefont {Haegeman}, \citenamefont {Lubich}, \citenamefont {Oseledets}, \citenamefont {Vandereycken},\ and\ \citenamefont {Verstraete}}]{haegeman2016unifying}%
  \BibitemOpen
  \bibfield  {author} {\bibinfo {author} {\bibfnamefont {J.}~\bibnamefont {Haegeman}}, \bibinfo {author} {\bibfnamefont {C.}~\bibnamefont {Lubich}}, \bibinfo {author} {\bibfnamefont {I.}~\bibnamefont {Oseledets}}, \bibinfo {author} {\bibfnamefont {B.}~\bibnamefont {Vandereycken}},\ and\ \bibinfo {author} {\bibfnamefont {F.}~\bibnamefont {Verstraete}},\ }\bibfield  {title} {\bibinfo {title} {Unifying time evolution and optimization with matrix product states},\ }\href@noop {} {\bibfield  {journal} {\bibinfo  {journal} {Phys. Rev. B}\ }\textbf {\bibinfo {volume} {94}},\ \bibinfo {pages} {165116} (\bibinfo {year} {2016})}\BibitemShut {NoStop}%
\bibitem [{\citenamefont {Frazier}(2018)}]{frazier2018tutorial}%
  \BibitemOpen
  \bibfield  {author} {\bibinfo {author} {\bibfnamefont {P.~I.}\ \bibnamefont {Frazier}},\ }\bibfield  {title} {\bibinfo {title} {A tutorial on {B}ayesian optimization},\ }\href@noop {} {\bibfield  {journal} {\bibinfo  {journal} {arXiv preprint arXiv:1807.02811}\ } (\bibinfo {year} {2018})}\BibitemShut {NoStop}%
\bibitem [{\citenamefont {Pedregosa}\ \emph {et~al.}(2011)\citenamefont {Pedregosa}, \citenamefont {Varoquaux}, \citenamefont {Gramfort}, \citenamefont {Michel}, \citenamefont {Thirion}, \citenamefont {Grisel}, \citenamefont {Blondel}, \citenamefont {Prettenhofer}, \citenamefont {Weiss}, \citenamefont {Dubourg}, \citenamefont {Vanderplas}, \citenamefont {Passos}, \citenamefont {Cournapeau}, \citenamefont {Brucher}, \citenamefont {Perrot},\ and\ \citenamefont {Duchesnay}}]{scikit-learn}%
  \BibitemOpen
  \bibfield  {author} {\bibinfo {author} {\bibfnamefont {F.}~\bibnamefont {Pedregosa}}, \bibinfo {author} {\bibfnamefont {G.}~\bibnamefont {Varoquaux}}, \bibinfo {author} {\bibfnamefont {A.}~\bibnamefont {Gramfort}}, \bibinfo {author} {\bibfnamefont {V.}~\bibnamefont {Michel}}, \bibinfo {author} {\bibfnamefont {B.}~\bibnamefont {Thirion}}, \bibinfo {author} {\bibfnamefont {O.}~\bibnamefont {Grisel}}, \bibinfo {author} {\bibfnamefont {M.}~\bibnamefont {Blondel}}, \bibinfo {author} {\bibfnamefont {P.}~\bibnamefont {Prettenhofer}}, \bibinfo {author} {\bibfnamefont {R.}~\bibnamefont {Weiss}}, \bibinfo {author} {\bibfnamefont {V.}~\bibnamefont {Dubourg}}, \bibinfo {author} {\bibfnamefont {J.}~\bibnamefont {Vanderplas}}, \bibinfo {author} {\bibfnamefont {A.}~\bibnamefont {Passos}}, \bibinfo {author} {\bibfnamefont {D.}~\bibnamefont {Cournapeau}}, \bibinfo {author} {\bibfnamefont {M.}~\bibnamefont {Brucher}}, \bibinfo {author} {\bibfnamefont {M.}~\bibnamefont {Perrot}},\ and\ \bibinfo {author} {\bibfnamefont
  {E.}~\bibnamefont {Duchesnay}},\ }\bibfield  {title} {\bibinfo {title} {Scikit-learn: Machine learning in {P}ython},\ }\href@noop {} {\bibfield  {journal} {\bibinfo  {journal} {J. Mach. Learn. Res.}\ }\textbf {\bibinfo {volume} {12}},\ \bibinfo {pages} {2825} (\bibinfo {year} {2011})}\BibitemShut {NoStop}%
\bibitem [{\citenamefont {Allahverdyan}\ \emph {et~al.}(2004)\citenamefont {Allahverdyan}, \citenamefont {Balian},\ and\ \citenamefont {Nieuwenhuizen}}]{allahverdyan}%
  \BibitemOpen
  \bibfield  {author} {\bibinfo {author} {\bibfnamefont {A.~E.}\ \bibnamefont {Allahverdyan}}, \bibinfo {author} {\bibfnamefont {R.}~\bibnamefont {Balian}},\ and\ \bibinfo {author} {\bibfnamefont {T.~M.}\ \bibnamefont {Nieuwenhuizen}},\ }\bibfield  {title} {\bibinfo {title} {Maximal work extraction from finite quantum systems},\ }\href@noop {} {\bibfield  {journal} {\bibinfo  {journal} {Europhys. Lett.}\ }\textbf {\bibinfo {volume} {67}},\ \bibinfo {pages} {565} (\bibinfo {year} {2004})}\BibitemShut {NoStop}%
\bibitem [{\citenamefont {Castellano}\ \emph {et~al.}(2024)\citenamefont {Castellano}, \citenamefont {Farina}, \citenamefont {Giovannetti},\ and\ \citenamefont {Acin}}]{castellano2024ELE}%
  \BibitemOpen
  \bibfield  {author} {\bibinfo {author} {\bibfnamefont {R.}~\bibnamefont {Castellano}}, \bibinfo {author} {\bibfnamefont {D.}~\bibnamefont {Farina}}, \bibinfo {author} {\bibfnamefont {V.}~\bibnamefont {Giovannetti}},\ and\ \bibinfo {author} {\bibfnamefont {A.}~\bibnamefont {Acin}},\ }\bibfield  {title} {\bibinfo {title} {Extended local ergotropy},\ }\href {https://doi.org/10.1103/PhysRevLett.133.150402} {\bibfield  {journal} {\bibinfo  {journal} {Phys. Rev. Lett.}\ }\textbf {\bibinfo {volume} {133}},\ \bibinfo {pages} {150402} (\bibinfo {year} {2024})}\BibitemShut {NoStop}%
\end{thebibliography}%

\appendix
\onecolumngrid
\section*{End Matter}
\twocolumngrid
We detail some points relevant for the analysis of the disordered XXZ Heisenberg chain. In particular, we provide additional informations on the entanglement entropy for the chain partition $N/2,N/2$, 
on the scaling behaviour of physical quantities with chain size $N$, 
and on the quantum fluctuations of local ergotropy in the case when the subsystem $S$ consists of two spins. 

\section{Entanglement entropy for the partition $N/2,N/2$}\label{app:ent_entr}
Here we report the analysis on the time behaviour of the entanglement entropy in the case of the chain partitioned into two exact halves, $N/2,N/2$, to highlight the differences between the three phases: MBL, AL and ERG. 
\begin{figure}[H]
\includegraphics[width=1\columnwidth]{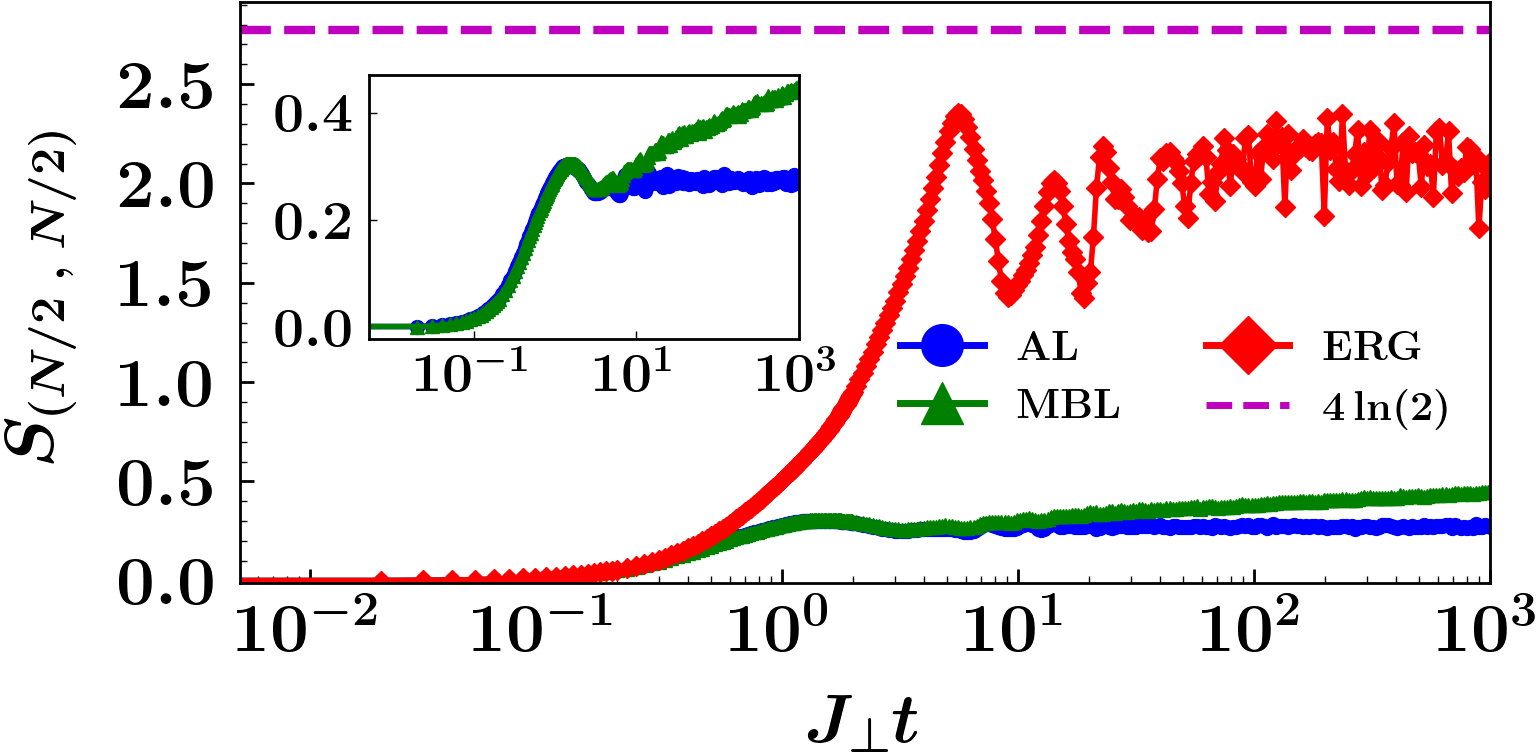}
\caption{\justifying Entanglement entropy for the $N/2,N/2$ partition, as a function of time in the three phases: AL (blue circle, $J_z/J_{\perp}=0$, $W/J_{\perp}=5$), MBL (green triangle, $J_z/J_{\perp}=0.2$, $W/J_{\perp}=5$), and ERG phase (red diamond, $J_z/J_{\perp}=0.2$, $W=0$). The dotted line indicates the saturation value of entanglement entropy for the system.  
The inset provides a zoomed-in view of the local ergotropy as a function of time in the two localized cases. Results are obtained by averaging over $10^3$ disorder realizations, with chain length $N=8$.}
\label{fig:ent_half}
\end{figure}
These results are qualitatively identical to those obtained in the $2,(N-2)$ case reported in Fig.\,\ref{fig:entanglement}, with only small differences at long times. In fact, considering a subsystem length of $N/2$, the finite-size saturation value for $N\ge 3$ is $S_{N/2,N/2}^{max}=(N/2)\ln 2 > S_{2,(N-2)}^{max}=2\ln 2$. 

In each of the three phases, the entropy starts from zero because of the initial Néel state. For a first transient of time, each phase exhibits a ballistic growth. After that, in the ergodic phase (red diamond in Fig.\,\ref{fig:ent_half}) entropy keeps growing towards the saturation value (dotted line in Fig.\,\ref{fig:ent_half}). In the two localized phases, the entropy growth halts around $J_{\perp}t\sim1$. In the MBL phase, entropy grows logarithmically, while, in the AL phase, it remains constant. 
These results are fully consistent with previous theoretical and experimental results.

\section{Dependence of physical quantities on size \texorpdfstring{$N$}{N}}\label{app:length}
Here we examine how several physical quantities scale with chain lengths $N$, as robustness to size increase is a key feature of MBL phenomenology \cite{Sierant}. 
\begin{figure}[H]
\includegraphics[width=1\columnwidth]{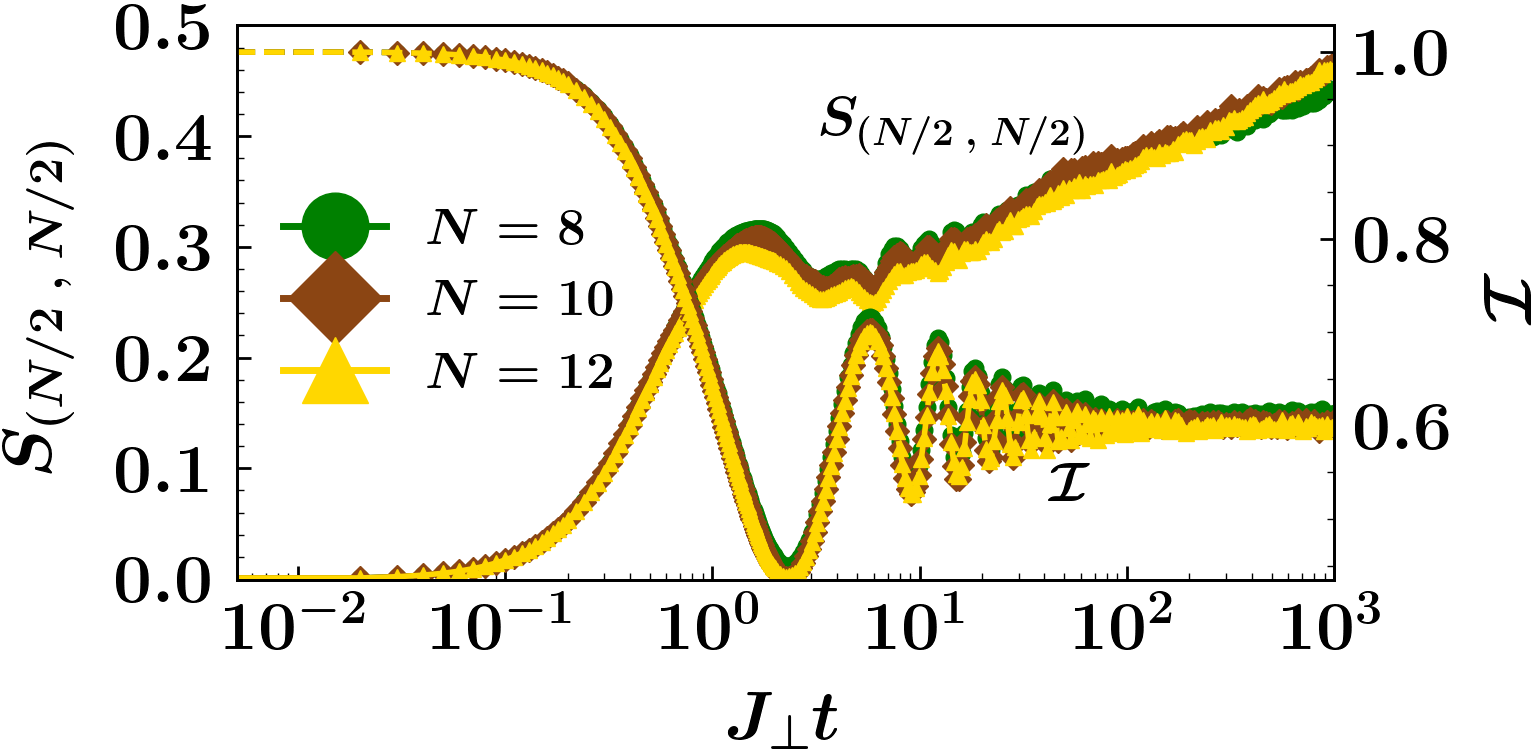}
  \caption{\justifying Entanglement entropy for $N/2,N/2$ partition, and imbalance $\mathcal{I}$ as functions of time for different sizes: $N=8$ (green circle), $N=10$ (brown diamond), and $N=12$ (gold triangle). Results are obtained by averaging over $1.6 \cdot 10^3$ disorder realizations, with coupling $J_z/J_{\perp}=0.1$, and disorder strength $W/J_{\perp}=5$.}
\label{fig:imbalance}
\end{figure}
In Fig.\,\ref{fig:imbalance}, we report the entanglement entropy, in the case of $N/2,N/2$ partition, as a function of time for different sizes. Moreover, we show the time behaviour of the imbalance defined as the normalized charge difference between odd and even sites, that in the spin representation reads: $\mathcal{I}=(\langle S^z_{o}\rangle-\langle S^z_{e}\rangle)/(\langle S^z_{o}\rangle+\langle S^z_{e}\rangle + N/2)$ (where sites first index is 1, see Refs. \cite{schreiber2015observation,Iemini2016}). From $N=8$ to $N=12$, the plots of both quantities show only minor differences.

\begin{figure}[]
  \begin{overpic}[width=1\columnwidth]{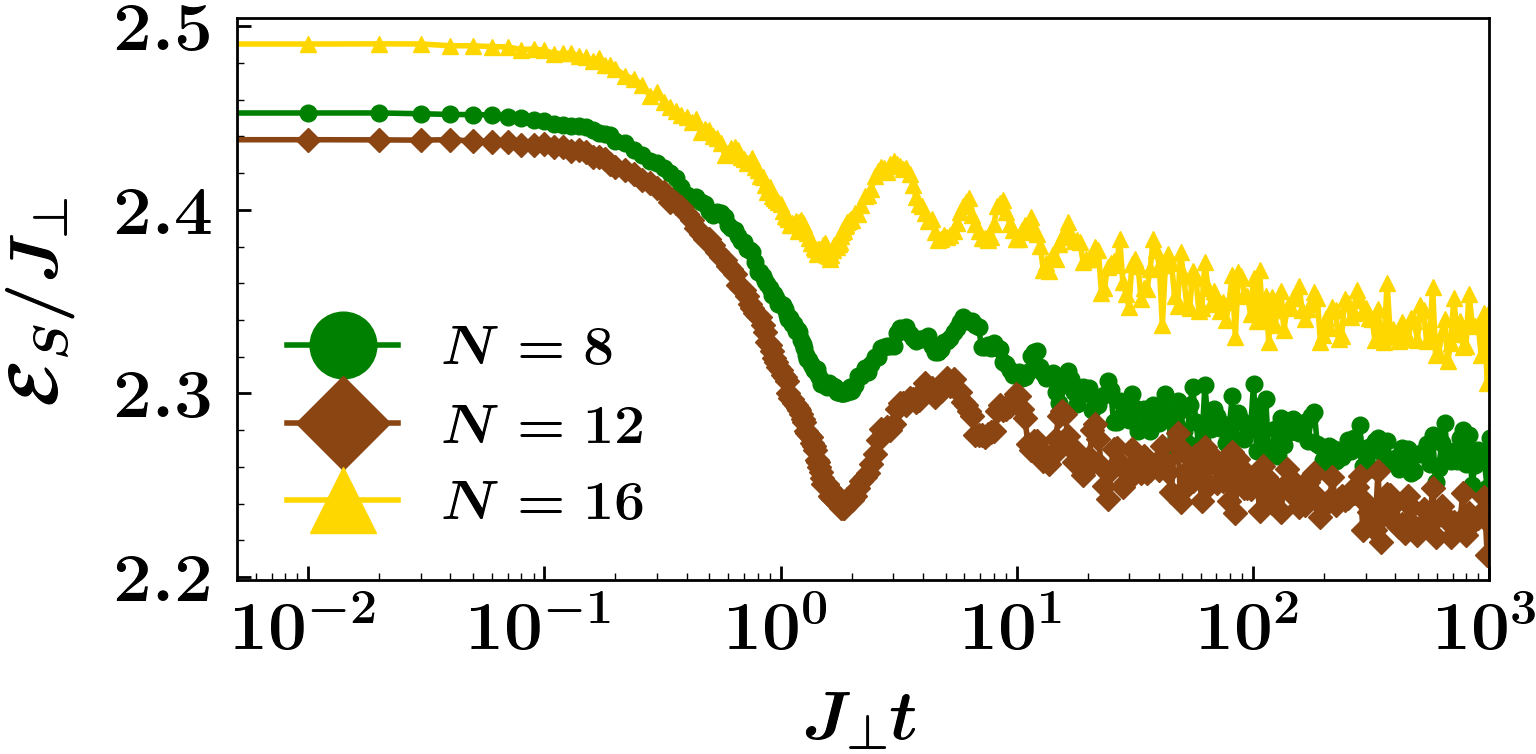}
    \put(-5.1, 43){\scalebox{1.5}{\textbf{(a)}}} 
  \end{overpic}
  \hspace{1.7cm}  
  \begin{overpic}[width=1\columnwidth]{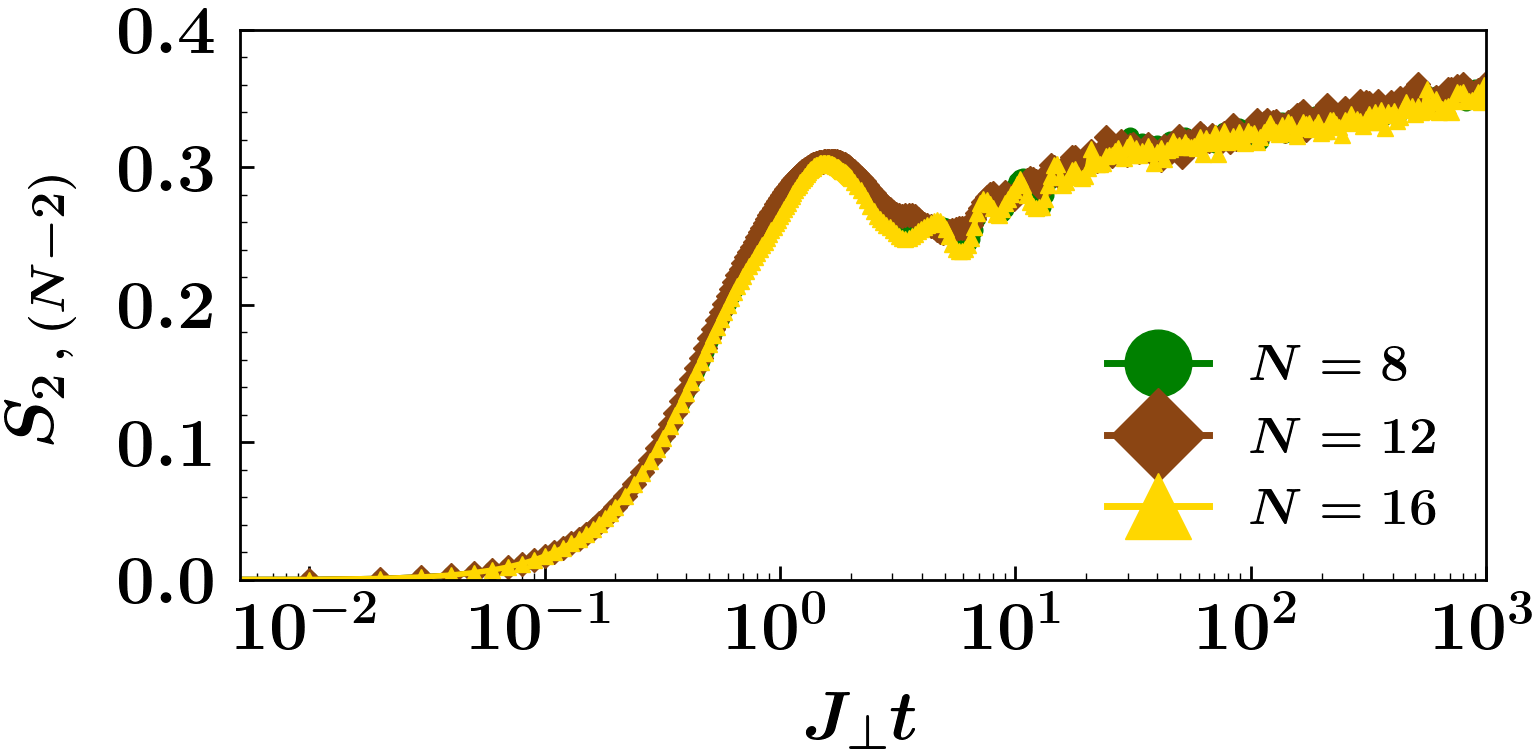}
   \put(-5.1, 46.6){\scalebox{1.5}{\textbf{(b)}}}  
  \end{overpic}
   \caption{\justifying (a) Local ergotropy and (b) entanglement entropy for $2,(N-2)$ partition as functions of time in the MBL phase ($J_z/J_{\perp}=0.2$, $W/J_{\perp}=5$) for various lengths: $N=8$ (green circle), $N=12$ (brown diamond) and $N=16$ (gold triangle). Results are obtained by averaging over $10^3$ disorder realizations for $N=8,12$ and around $6.6 \cdot 10^2$ for $N=16$.} 
  \label{fig:quantum_relative_fluctuations1}
\end{figure}

The plots in the upper panel of Fig.\,\ref{fig:quantum_relative_fluctuations1} reveal that local ergotropy changes slightly with increasing $N$ in the MBL phase. Only at intermediate values of time ($J_{\perp}t \simeq 1$), there is a deviation with varying $N$. This could be attributed to the fact that the local ergotropy is obtained also through a BOA. In the localized phases, $N=8$ is close to the thermodynamic limit within the parameter ranges analyzed in this work. However, in the ERG phases finite-size effects are more pronounced. Finally, in the lower panel of Fig.\,\ref{fig:quantum_relative_fluctuations1}, we report the entanglement entropy, in the case of $2,(N-2)$ chain partition, as a function of time for different sizes. In analogy with the entanglement entropy for the $N/2,N/2$ chain partition (see Fig.\,\ref{fig:imbalance}), very small differences are obtained with increasing $N$. 

\section{Quantum fluctuations}
\label{app:quantum_fluc}
\begin{figure}[H]
  \begin{overpic}[width=1\columnwidth]{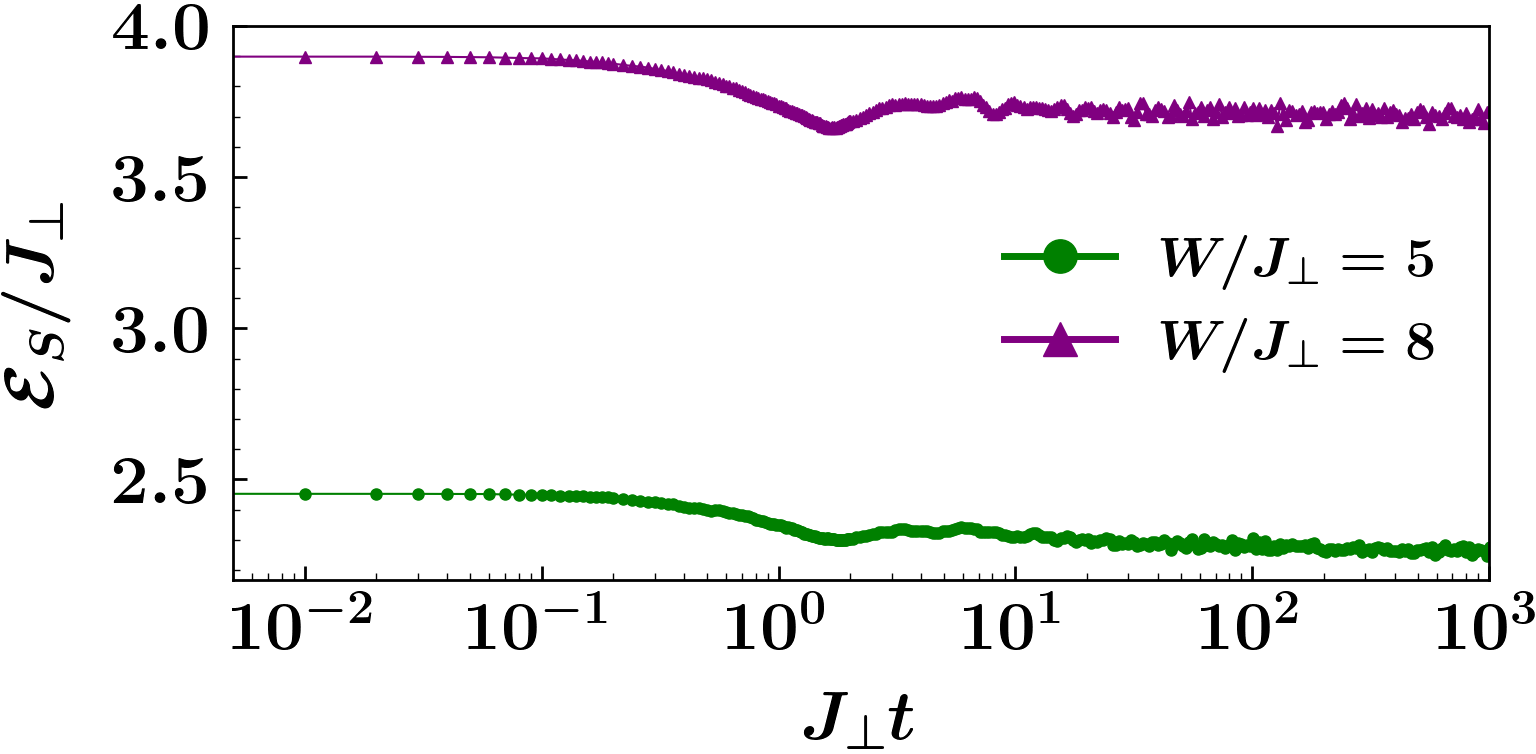}
    \put(-5.8, 43.6){\scalebox{1.5}{\textbf{(a)}}} 
  \end{overpic}
  \hspace{1.7cm}  
  \begin{overpic}[width=1\columnwidth]{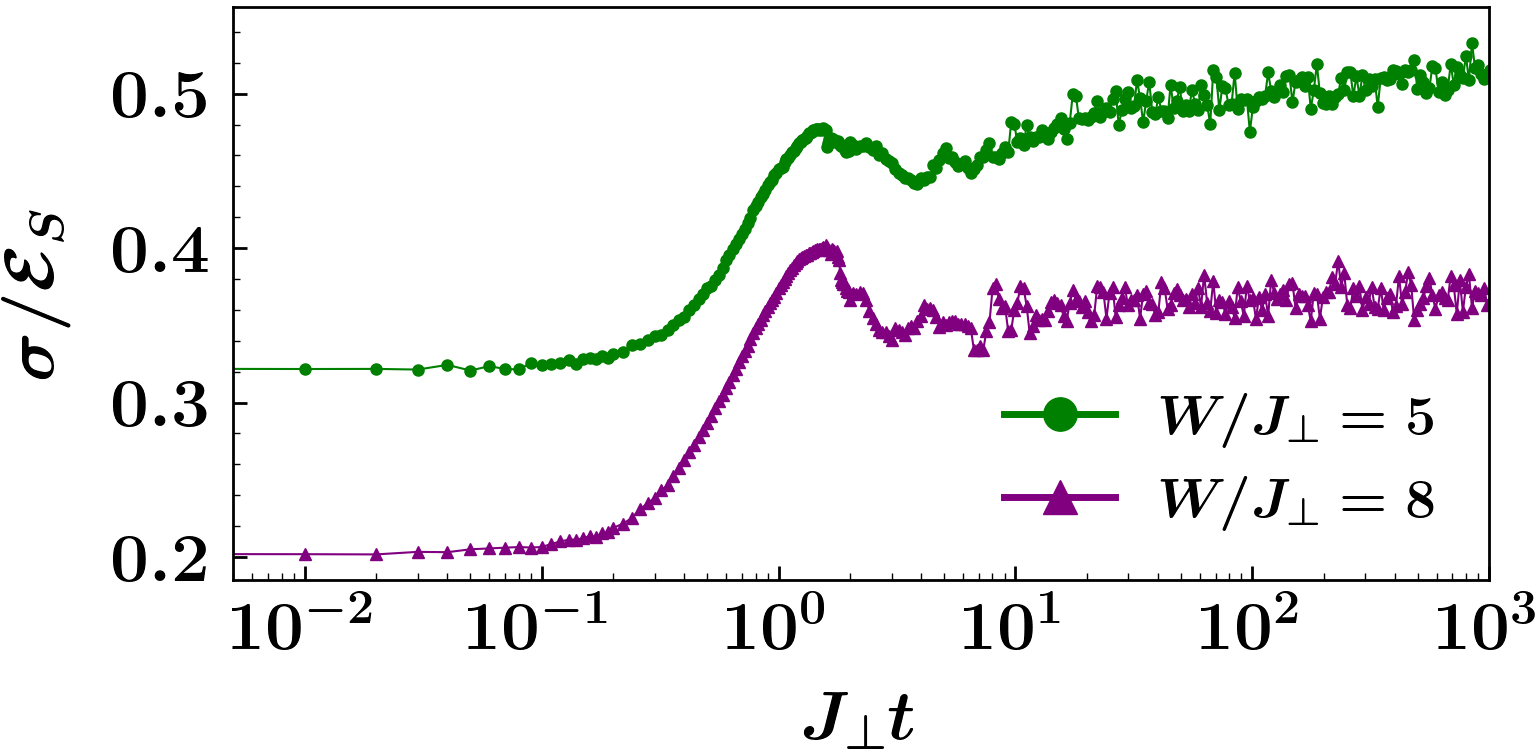}
   \put(-5.8, 46.6){\scalebox{1.5}{\textbf{(b)}}}  
  \end{overpic}
   \caption{\justifying (a) Local ergotropy and (b) its relative quantum fluctuations as functions of time for different values of the disorder strength $W$ in the MBL phase ($J_z/J_{\perp}=0.2$): $W/J_{\perp}=5.0$ (green circle), $W/J_{\perp}=8.0$ (magenta triangle). Results are obtained by averaging over $10^3$ disorder realizations, with chain length $N=8$.} 
  \label{fig:quantum_relative_fluctuations}
\end{figure}

We report here details relative to quantum fluctuations of local ergotropy additional to those presented in Fig.\,\ref{fig:fluttuazioni_quantistiche}. In Fig.\,\ref{fig:quantum_relative_fluctuations} (a)
we plot the local ergotropy with increasing disorder strength $W$. The local ergotropy depends mostly on $W$ for large disorder $W/J_{\perp} \gg 1$ \cite{suppmat}. 
More precisely, it is of the order of $W/2$. Thus, as disorder strength increases, local ergotropy and its quantum fluctuations also increase. However, as shown in Fig.\,\ref{fig:quantum_relative_fluctuations} (b) the relative quantum fluctuations decrease with increasing $W$, underscoring the robustness of localization.

\end{document}


\begin{center}
\textbf{\Large Supplemental Material for\\ \enquote{Local ergotropy dynamically witnesses many-body localized phases}}
\end{center}

\twocolumngrid
We provide additional details on fluctuations over disorder realizations, 
 on the dependence of the local ergotropy on the spin-spin interaction strength $J_z$, 
 on the order of magnitude of the local ergotropy, 
 on the global ergotropy, 
 on Jordan-Wigner transformations, 
and on the numerical methods. 
As in the main text, we analyze the disordered XXZ Heisenberg chain.       

\section{Fluctuations over disorder realizations}\label{sec:Disorder}

In addition to the quantum fluctuations of the local ergotropy, we have also computed its fluctuations over disorder realizations. One expects that the small oscillations shown by the numerical results of local ergotropy at long times arise not only from finite-size effects but also from averaging over a finite large number of disorder realizations. 

To evaluate the fluctuations for a single local ergotropy realization at time $t$, one can use the standard deviation
\begin{equation}
\sigma_{\epsilon}^{cl}(t)= \sqrt{   \frac{1}{R-1}\sum\limits_{i=1}^R ( \mathcal{E}_i(t)-\overline{\mathcal{E}(t)})^2},
\end{equation}
where $R$ is the number of the disorder realizations over which we average, $\mathcal{E}_i(t)$ is the single ergotropy realization at time $t$, and $\overline{\mathcal{E}(t)}$ is the disorder-averaged local ergotropy at time $t$. 

To quantify the fluctuations for disorder-averaged ergotropy at time $t$, we can use the mean standard deviation:

\begin{equation}\label{eq:sigmacl}
\sigma^{cl}_{\overline{\epsilon}}(t)= \sigma_{\epsilon}^{cl}(t)/\sqrt{R}.
\end{equation}

\begin{figure}[htbp!]
  \begin{overpic}[width=1\columnwidth]{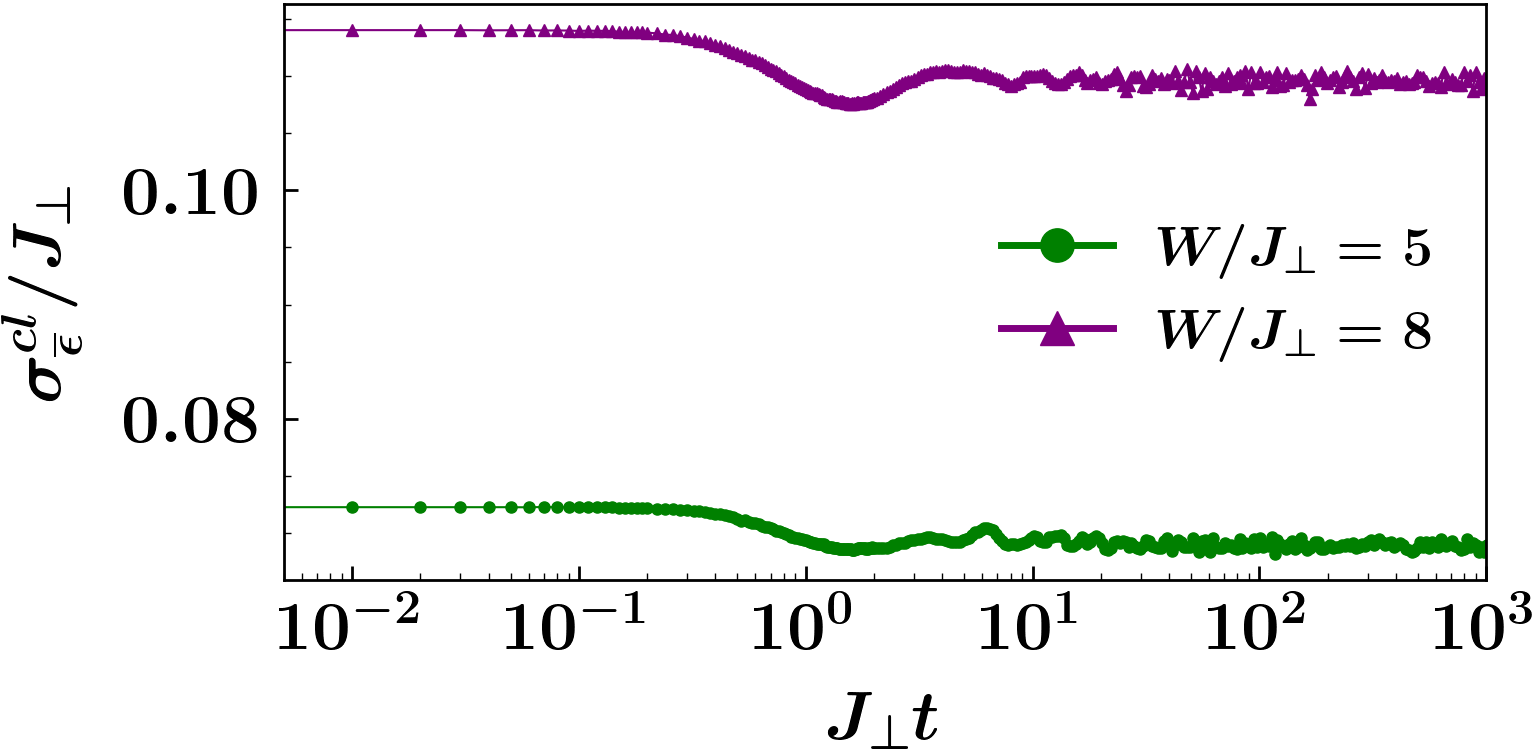}
    \put(-4.5, 46.6){\scalebox{1.5}{\textbf{(a)}}} 
  \end{overpic}
  \hspace{1.7cm}  
  \begin{overpic}[width=1\columnwidth]{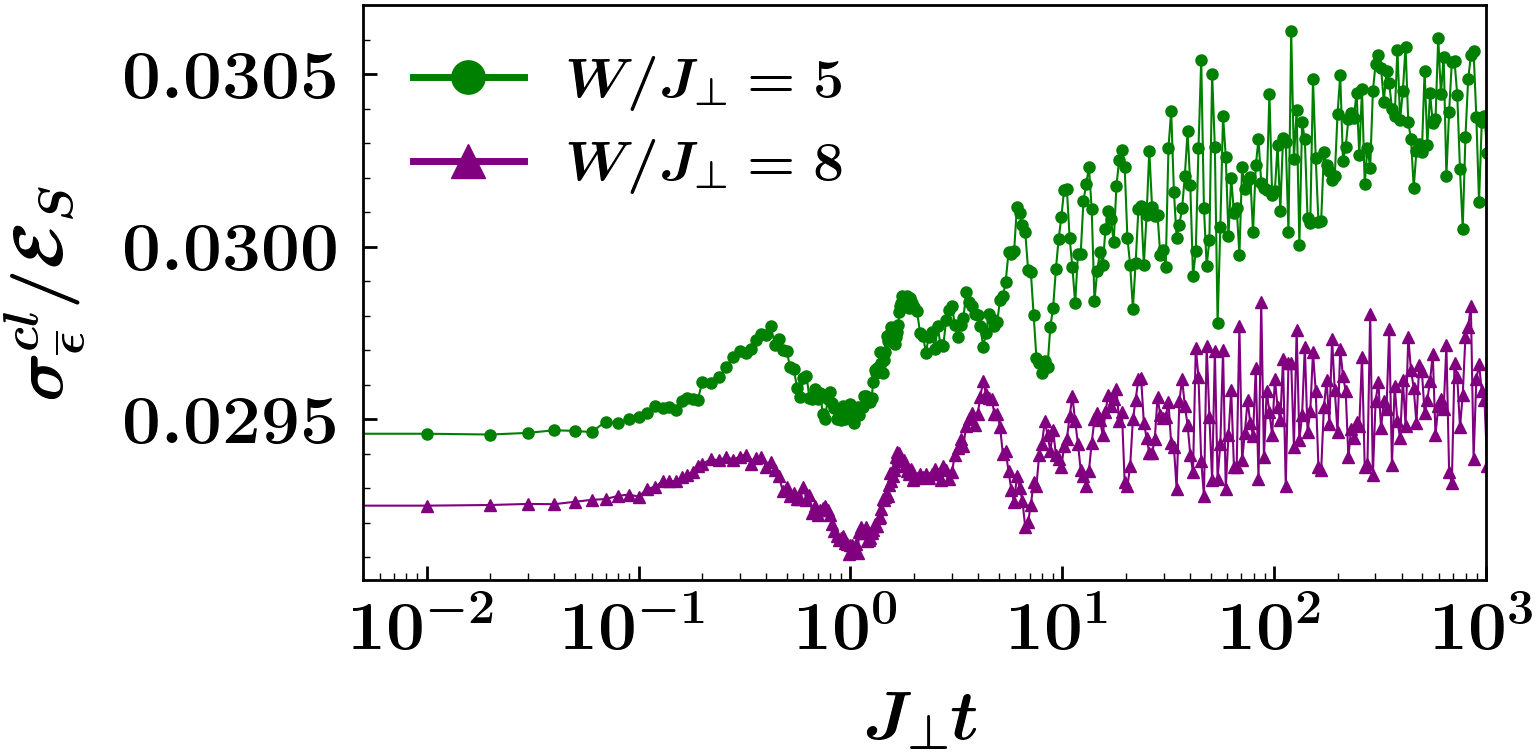}
   \put(-4.5, 46.6){\scalebox{1.5}{\textbf{(b)}}}  
  \end{overpic}
   \caption{\justifying (a) Fluctuations and (b) relative fluctuations of local ergotropy over disorder realizations as functions of time for different values of the disorder strength $W$ in the MBL phase ($J_z/J_{\perp}=0.2$): $W/J_{\perp}=5.0$ (green circle), $W/J_{\perp}=8.0$ (purple triangle). Results are obtained by averaging over $10^3$ disorder realizations, with chain length of $N=8$.} 
  \label{fig:fluttua_classiche}
\end{figure}

Fig.\,\ref{fig:fluttua_classiche} reveals that the disorder-induced fluctuations are small in comparison with the quantum fluctuations discussed in the main text. As expected, the upper panel of Fig.\,\ref{fig:fluttua_classiche} shows that disorder fluctuations increase with the disorder strength. In fact, their order of magnitude, estimated via the mean standard deviation, scales as $(W/R)^{1/2}$. In contrast to quantum fluctuations, the lower panel of Fig.\,\ref{fig:fluttua_classiche} points out that the relative fluctuations due to disorder are almost constant as the disorder strength increases.

\section{Dependence of local ergotropy on the interaction strength \texorpdfstring{$J_z$}{J_z}}\label{sec:interaction}


Here we report the analysis conducted for the time behaviour of the local ergotropy as a function of longitudinal interaction strength $J_z$. 

\begin{figure}[H]
  \begin{overpic}[width=1\columnwidth]{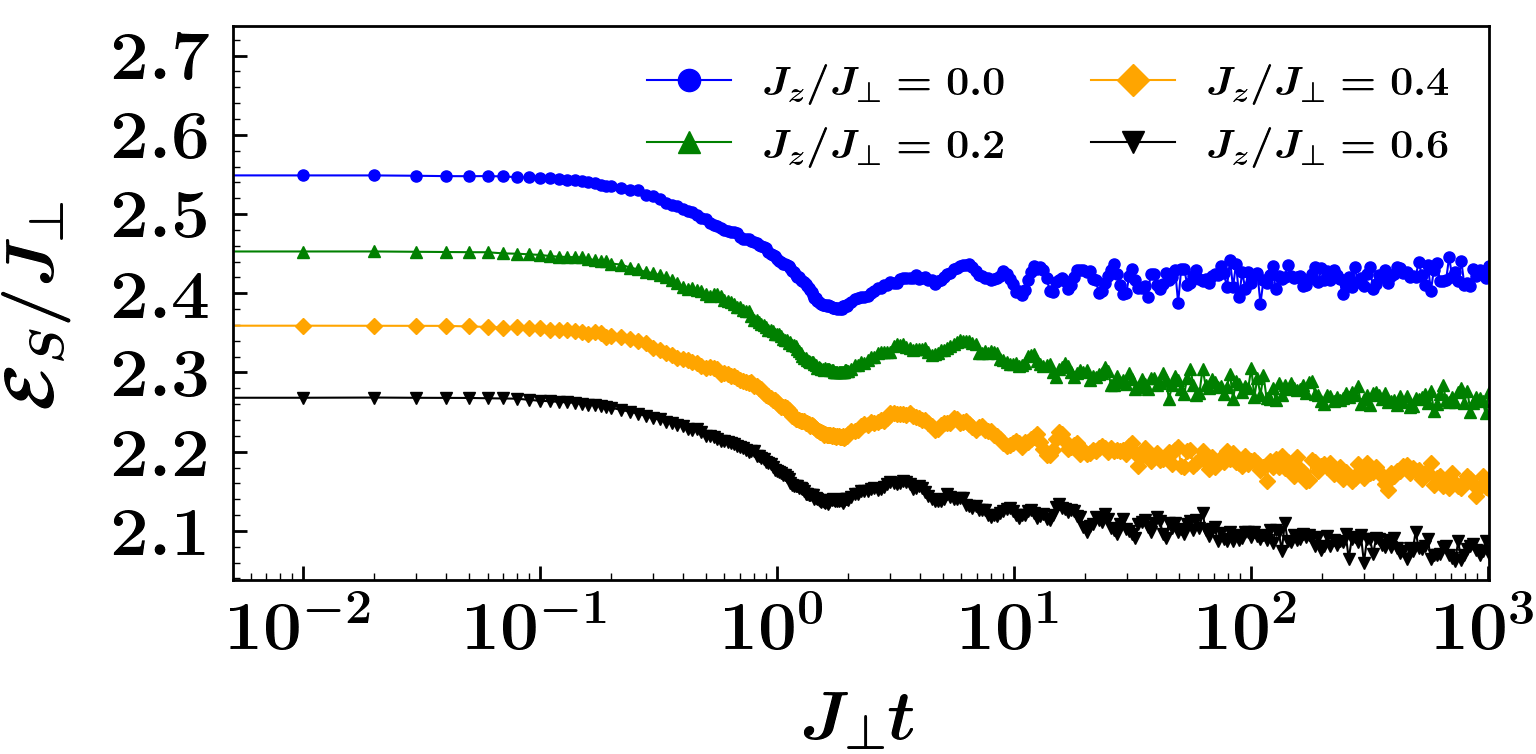}
    \put(-5.8, 46.6){\scalebox{1.5}{\textbf{(a)}}} 
  \end{overpic}
  \hspace{1.7cm}  
  \begin{overpic}[width=1\columnwidth]{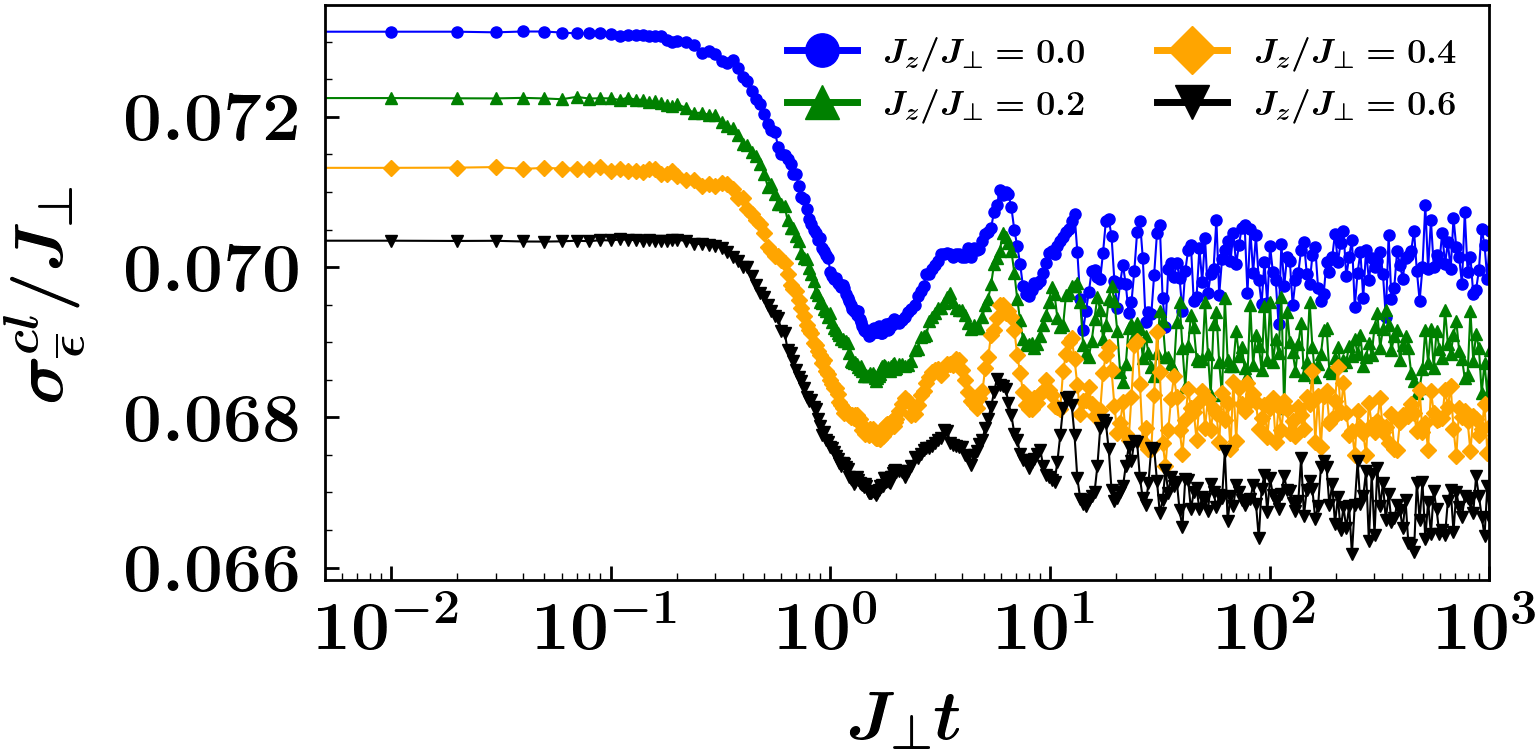}
   \put(-5.8, 46.6){\scalebox{1.5}{\textbf{(b)}}}  
  \end{overpic}
   \caption{\justifying (a) Local ergotropy and (b) its fluctuations over disorder realizations as functions of time at $W/J_{\perp}=5.0$ for different values of the longitudinal interaction strength $J_z$: AL phase for $J_z/J_{\perp}=0$ (blue circle), MBL phase for $J_z/J_{\perp}=0.2$ (green upward triangle), $J_z/J_{\perp}=0.4$ (orange diamond), $J_z/J_{\perp}=0.6$ (black downward triangle). Results are obtained by averaging over $10^3$ disorder realizations, with chain length of $N=8$.} 
  \label{fig:dev_st_Jz}
\end{figure}

From the upper panel of Fig.\,\ref{fig:dev_st_Jz}, we observe that, for the considered parameter range, lower values of local ergotropy correspond to higher values of $J_z$. In fact, stronger interactions lead to more energy being trapped in the interaction terms, making less energy available to be extracted as work \cite{Polini}. As discussed in the main text, AL phase (corresponding to $J_z=0$) shows the highest ergotropy. However, the lower panel of Fig.\,\ref{fig:dev_st_Jz} points out that the fluctuations over disorder realizations decrease, as $J_z$ increases. Therefore, over the disorder realizations, the local ergotropy is more fluctuating in the AL than in MBL phase, in agreement with results reported in the literature \cite{Polini}. We remark that this behaviour is opposite to that found for quantum fluctuations. 

\section{Order of magnitude of local ergotropy}\label{sec:order-of-magnitude}

\begin{figure}[H]
  \includegraphics[width=1\columnwidth]{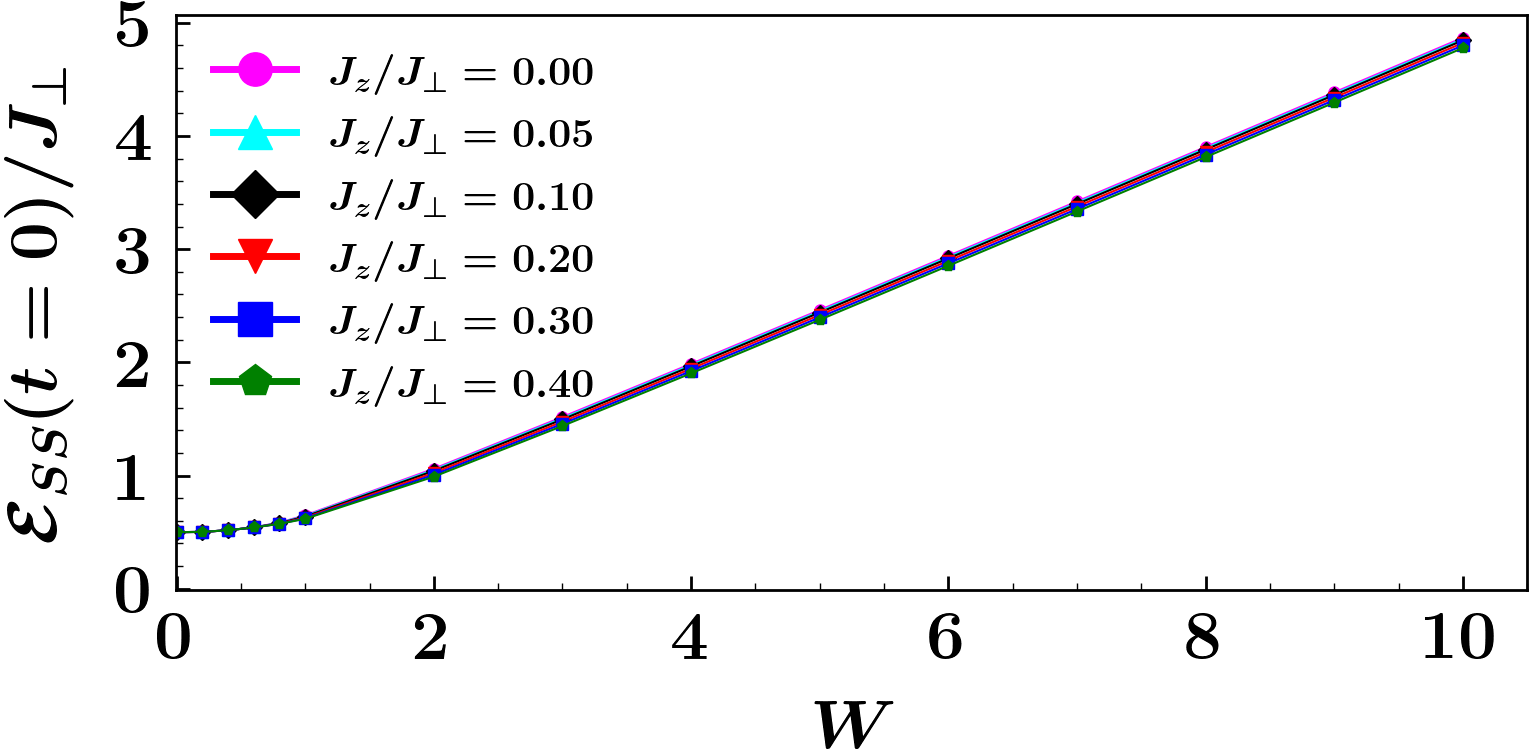}
  \caption{\justifying Subsystem ergotropy at $t=0$ as a function of the disorder strength $W$, for various values of $J_z$. Results are obtained by averaging over $250$ disorder realizations, with chain length of $N=8$. }
\label{fig:va_come}
\end{figure}

Since the considered values of the interaction strength $J_z$ are small compared to the disorder strength $W$, the order of magnitude of the local ergotropy is expected to be dominated by $W$. In fact, for large $W$, the ergotropy value per site is of the order of $W/4$. Therefore, for a subsystem made of two spins, the local ergotropy scales as $W/2$.   

The previous estimate comes from the dependence on the disorder strength $W$ of the subsystem ergotropy evaluated at the initial time $t=0$. 
As shown in Fig.\,\ref{fig:va_come}, for $W >1$, the dependence of the subsystem ergotropy on $W$ is linear, with a slope of $1/2$. Since localized systems preserve memory of the initial state indefinitely, the time-evolved quantum state remains close to the Néel state for the time scales under consideration. Even if the above calculation and plot regard solely the initial time $t=0$, the local ergotropy is expected to maintain the same order of magnitude also for $t > 0$, as confirmed by the local ergotropy in Fig.2 in the main text.

\section{Global ergotropy}\label{sec:Global}
\begin{figure}[H]
  \begin{overpic}[width=1\columnwidth]{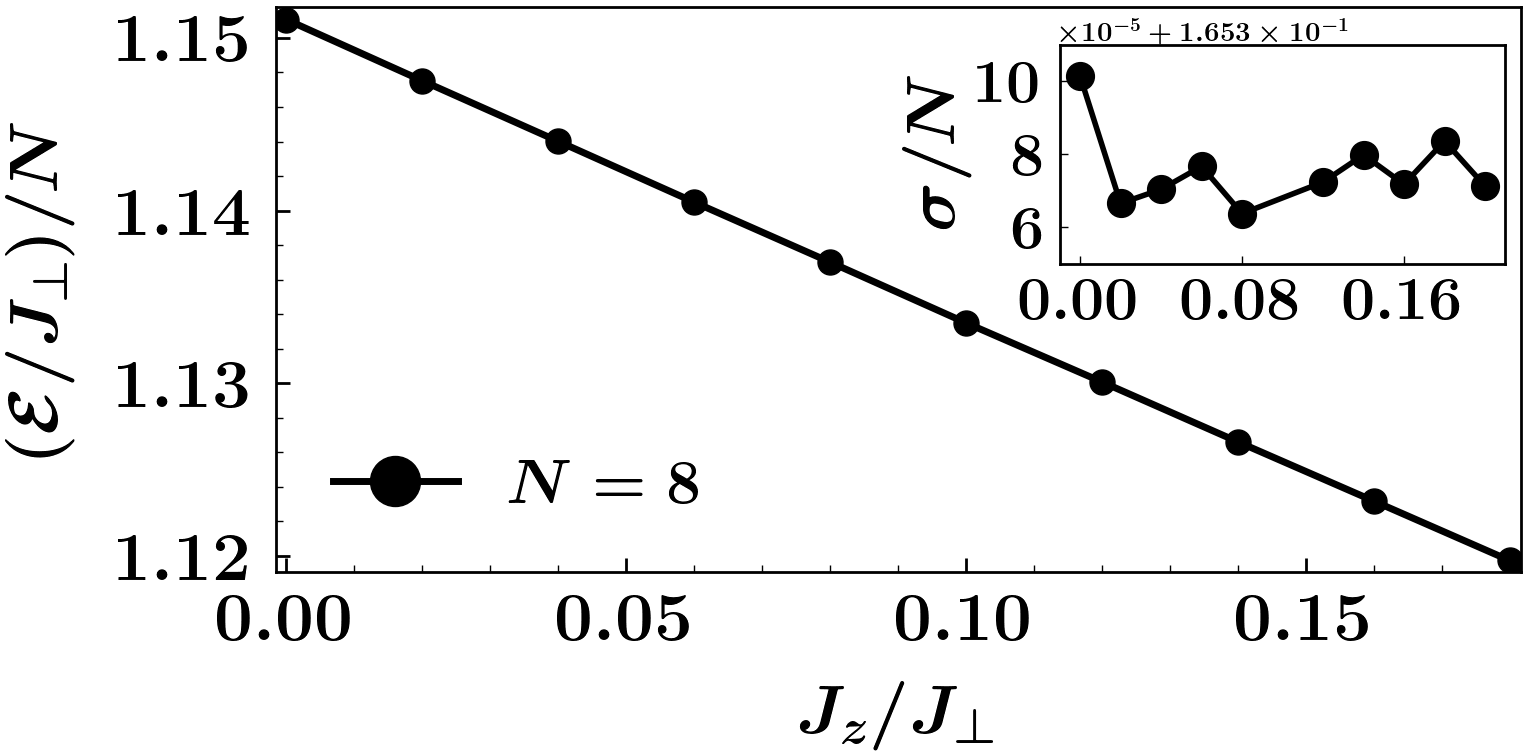}
    \put(-3.8, 46.6){\scalebox{1.5}{\textbf{(a)}}} 
  \end{overpic}
  \hspace{1.7cm}  
  \begin{overpic}[width=1\columnwidth]{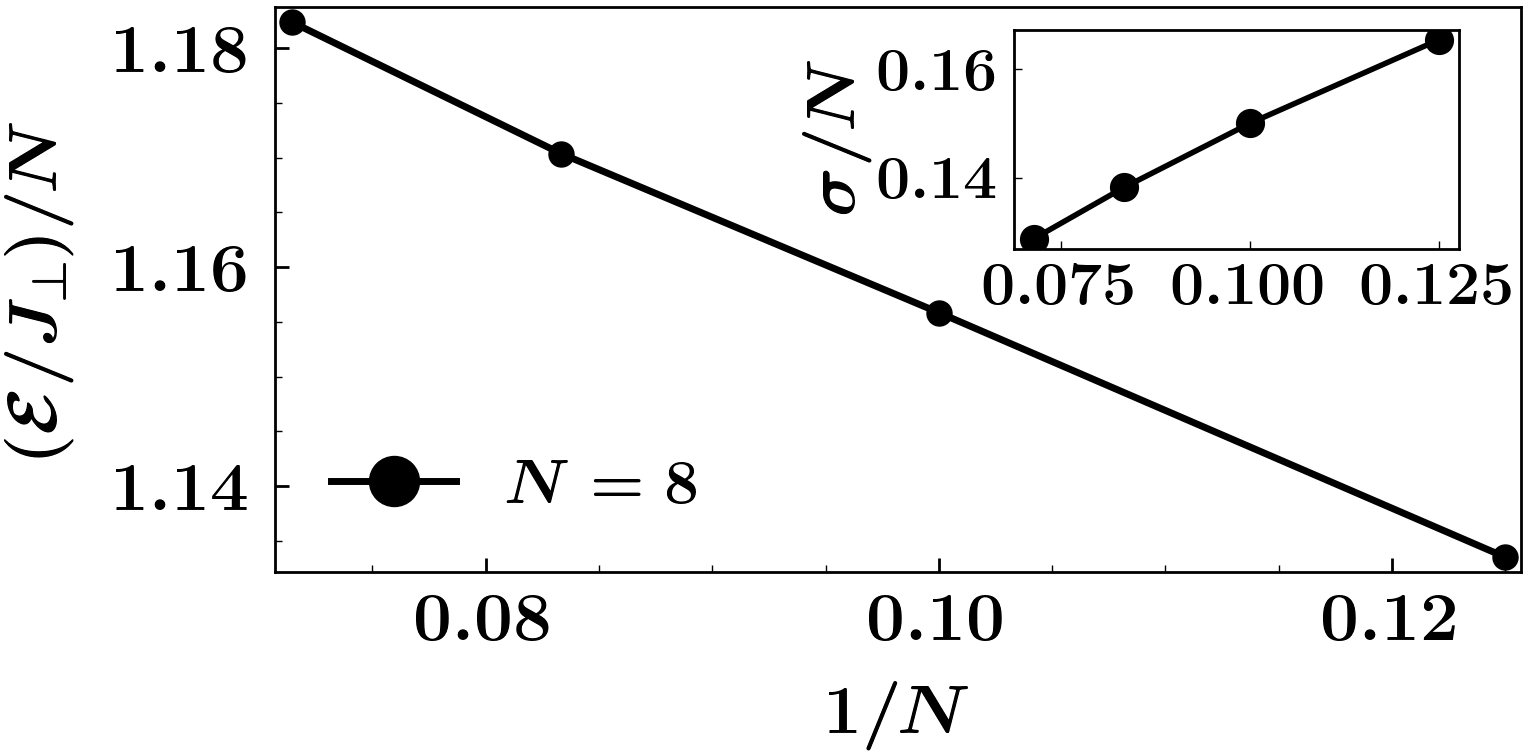}
   \put(-3.8, 46.6){\scalebox{1.5}{\textbf{(b)}}}  
  \end{overpic}
   \caption{\justifying (a) Global ergotropy per site for the $SE$ compound as a function of the interaction strength $J_z/J_{\perp}$. The inset shows quantum fluctuations per site. Results are obtained for $N=8$ and by averaging over $2.8 \cdot 10^5$ disorder realizations. (b) Global ergotropy per site for the $SE$ compound as a function of the inverse of the chain length, $1/N$. The inset shows quantum fluctuations per site. Results are obtained for $J_z/J_{\perp}=0.1$ and by averaging over $2.8 \cdot 10^5$ disorder realizations.} 
  \label{fig:global_ergo}
\end{figure}

We performed extensive simulations of the global ergotropy $\mathcal{E}$ for various values of interaction strength $J_z$ and chain length $N$. Since the global ergotropy is the disorder-averaged difference between the mean energy of the initial charged state (the Néel state) and the mean energy of the final state (the ground state) \cite{campaioli2024colloquium} its calculation requires only the use of the density matrix renormalization group (DMRG) to compute the ground state, without the need of time evolution. So we have pushed the simulations up to $2.8 \cdot 10^4$ disorder realizations. 

This analysis has two main purposes. First, it provides a benchmark for the order of magnitude of the local ergotropy, since, according to the theory, the global ergotropy is an upper bound for the local one. Second, since the Néel state is not an eigenstate of the XXZ Heisenberg model (even in the absence of disorder), it exhibits intrinsic fluctuations. It is important to verify that these fluctuations are smaller than the ergotropy value. Our results, as shown in Fig.\,\ref{fig:global_ergo} (a), have confirmed that these fluctuations are one order of magnitude smaller than ergotropy.  

Furthermore, Fig.\,\ref{fig:global_ergo} (b) shows that the global ergotropy exhibits a nearly linear dependence on $1/N$. This is important to understand how this quantity reaches the thermodynamic limit.

\section{Jordan-Wigner transformations}\label{sec:Jordan-Wigner}
The starting model, on which the present study is based, is a one-dimensional disordered electron model with nearest-neighbour particle-particle interactions, hard-wall boundary conditions, and infinite on-site Coulomb repulsion, ensuring that the electrons are spinless: 
\begin{equation}\label{eq:hsyst}
\hat{H}= -t \sum\limits^{{N-1}}_{i=1}
    \bigg\{\hat{c}^{\dag}_i \hat{c}_{i+1} + \hat{c}^{\dag}_{i+1} \hat{c}_i \bigg\} +  {V} \sum\limits^{N-1}_{i=1} \hat{n}_{i}\hspace{0.5mm}\hat{n}_{i+1}+\sum\limits^{N}_{i=1}H_i \hspace{1mm}\hat{n}_{i},
\end{equation}
where $t$ is the hopping energy, $N$ is the number of chain sites, $V$ denotes the interaction energy, while $H_i$ accounts for the disorder. The operator $ \hat{c}_i $  ($\hat{c}^{\dag}_i$) annihilates (creates) an electron at the site $i$. The fermionic system is assumed for simplicity in half-filling conditions.    

Jordan-Wigner transformations establish a mapping between quantum one-dimensional fermionic and spin systems. Indeed, the “down” and “up” states of a single spin correspond to empty and singly occupied fermion states. This allows us to write:
\begin{equation}
    \begin{cases}
\ket{\uparrow}\coloneqq\hat{c}^{\dag}\ket{0}\to\hat{S}^{(+)}\coloneqq\hat{c}^{\dag},
    \hspace{2cm}
    \\[2.6ex]
  \ket{\downarrow}\coloneqq\ket{0}\to\hat{S}^{(-)}\coloneqq\hat{c},
\\[2.6ex]
 \hat{S}^{(z)}\coloneqq\hat{n}-\frac{1}{2}=\hat{c}^{\dag}\hat{c}-\frac{1}{2},
 \\[1.7ex]
       \hat{S}^{(x)}=\frac{1}{2}( \hat{S}^{(+)} + \hat{S}^{(-)})=\frac{1}{2}(\hat{c}^{\dag}+\hat{c}),
    \\[1.7ex]
    \hat{S}^{(y)}=\frac{1}{2\mathrm{i}}( \hat{S}^{(+)} - \hat{S}^{(-)})=\frac{1}{2\mathrm{i}}(\hat{c}^{\dag}-\hat{c}).
    \end{cases}
\end{equation}
For multiple spins, spin operators commute on different sites, while fermionic ones anticommute. To maintain consistency, this representation needs to be modified:
\begin{equation}
    \begin{cases}
\hat{S}_j^{(z)}=\hat{c}_j^{\dag}\hat{c}_j-\frac{1}{2}
        \\[1.7ex]
\hat{S}_j^{(+)}=e^{\mathrm{i}\pi \sum\limits_{l<j}\hat{c}_l^{\dag}\hat{c}_l}\hat{c}_j^{\dag}=\hat{c}_j^{\dag}e^{-\mathrm{i}\pi \sum\limits_{l<j}\hat{c}_l^{\dag}\hat{c}_l}
        \\[1.7ex]
        \hat{S}_j^{(-)}=e^{\mathrm{i}\pi \sum\limits_{l<j}\hat{c}_l^{\dag}\hat{c}_l}\hat{c}_j=\hat{c}_je^{-\mathrm{i}\pi \sum\limits_{l<j}\hat{c}_l^{\dag}\hat{c}_l}.
    \end{cases}
\end{equation}

Using the Jordan-Wigner transformations, the fermionic system can be mapped onto the one-dimensional disordered XXZ Heisenberg model with nearest-neighbour interactions given in Eq. (2) of main text. Hard-wall boundary conditions are imposed in the subspace $S^{(z)}_{tot}=0$, being $S^{(z)}$ the z-component of the total spin. In Eq. (2) of main text, $J_{\perp}=2t$ is the spin-flip transverse interaction energy, $J_z=V$ is the longitudinal interaction energy, and $h_i=H_i$ are the on-site energy terms associated with an external magnetic field.


\section{Numerical methods}\label{sec:numerical-methods}
Since the model is disordered, we perform averages over different realizations of the Hamiltonian in Eq. (2) of main text. Every simulation starts from a different configuration of disorder, initialized through a stochastic extraction of the magnetic field values at each site. 

We adopt the tensor networks coding to perform the calculations, representing the quantum states in the matrix product state (MPS) form \cite{perez2006matrix}, and the operators in the matrix product operator (MPO) form \cite{white1992density,schollwock2011density}. The density matrix renormalization group (DMRG) is employed to compute the ground state of the $SE$ compound and the two-site time dependent variational principle (TDVP) to evolve the quantum state \cite{haegeman2011time,haegeman2016unifying}. 

In every simulation, the system starts in the Néel state. The TDVP is used to compute the time-evolved charged quantum state. At each time step, we compute all the quantities we need. Specifically, we compute the energy $\expval{\hat{H}}=\matrixel{\psi(t)}{\hat{H}}{\psi(t)}$,  $\expval{\hat{H^2}}$ to calculate the quantum fluctuations, and the subsystem reduced density matrix $\hat{\rho}_S(t)$ for the entanglement entropy.

To evaluate the time-dependent local ergotropy, we need to compute at each time step the maximum work extractable by applying an optimal unitary transformation to the charged state. However, the unitary operator is unknown, and changes from time to time. Hence, we use a Bayesian optimization algorithm (BOA) with a Gaussian process regression implemented through Python scikit-learn to find the best unitary operator \cite{frazier2018tutorial, scikit-learn}. 

Given the transformation to be used for local ergotropy 
\begin{equation}
    \hat{U}(t)= \hat{U}_S(t) \otimes \hat{I}_E,
\end{equation}
we can write in full generality a decomposition of the local unitary operator as follows: 
\begin{equation}
    \hat{U}_S(t)=\hat{U}_{AL}(t) \hspace{0.5mm}\cdot\hat{U}_1(t).
\end{equation}
The first term, $\hat{U}_{AL}$, is the optimal evolution operator defined in \cite{allahverdyan} for the isolated subsystem. It is constructed using the spectral decompositions of $\hat{H}_S$ and $\hat{\rho}_S(t)$:
\begin{equation}
\begin{split}
    \hat{\rho}_S(t)=\sum\limits_i r_i\ket{r_i}\bra{r_i},& \hspace{1cm}   \hat{H}_S=\sum\limits_k \epsilon_k\ket{\epsilon_k}\bra{\epsilon_k}
    \\
    r_i\geq r_{i+1},& \hspace{1cm}  \epsilon_k\leq \epsilon_{k+1}.
\end{split}
\end{equation}
The corresponding unitary operator is: 
\begin{equation}
    \hat{U}_{AL}=\sum\limits_j \ket{\epsilon_j}\bra{r_j}.
\end{equation}
This operator is the optimal one for the subsystem ergotropy $\mathcal{E}_{SS}$ and can be used to find a lower bound for the local ergotropy $\mathcal{E}_{S}$ \cite{castellano2024ELE}.

The second operator, $\hat{U}_1$, is not known a priori. In order to minimize the final energy over all local operators acting on the subsystem $S$, we require a parametrized representation of this operator, which is optimized to maximize the extractable work at each selected time step. Since our model is made of spin $1/2$ particles, the unitary operator is constructed as an exponential of a Hermitian operator $\hat{A}$, which can be written as linear combination of all the Pauli matrices $\hat{\sigma}^i\in\{\hat{I}, \hat{\sigma}^x, \hat{\sigma}^y, \hat{\sigma}^z \}$ (including the identity $\hat{I}$):
\begin{equation}
   \begin{split}
    \hat{U}_1=e^{-\mathrm{i}\hat{A}}, 
    \hspace{5mm}
    \\
    \hspace{5mm}\hat{A}=\sum\limits_{ij}a_{ij}  \hspace{1mm}\hat{\sigma}_{1}^i\otimes \hat{\sigma}_{2}^j, 
   \end{split}
\end{equation}
with $i,j=0,1,2,3$.
The 15 parameters $a_{ij}$ (excluding $i=j=0$ which adds an irrelevant phase) are optimized at each time step to maximize the ergotropy.

\begin{figure}[H]
  \begin{overpic}[width=1\columnwidth]{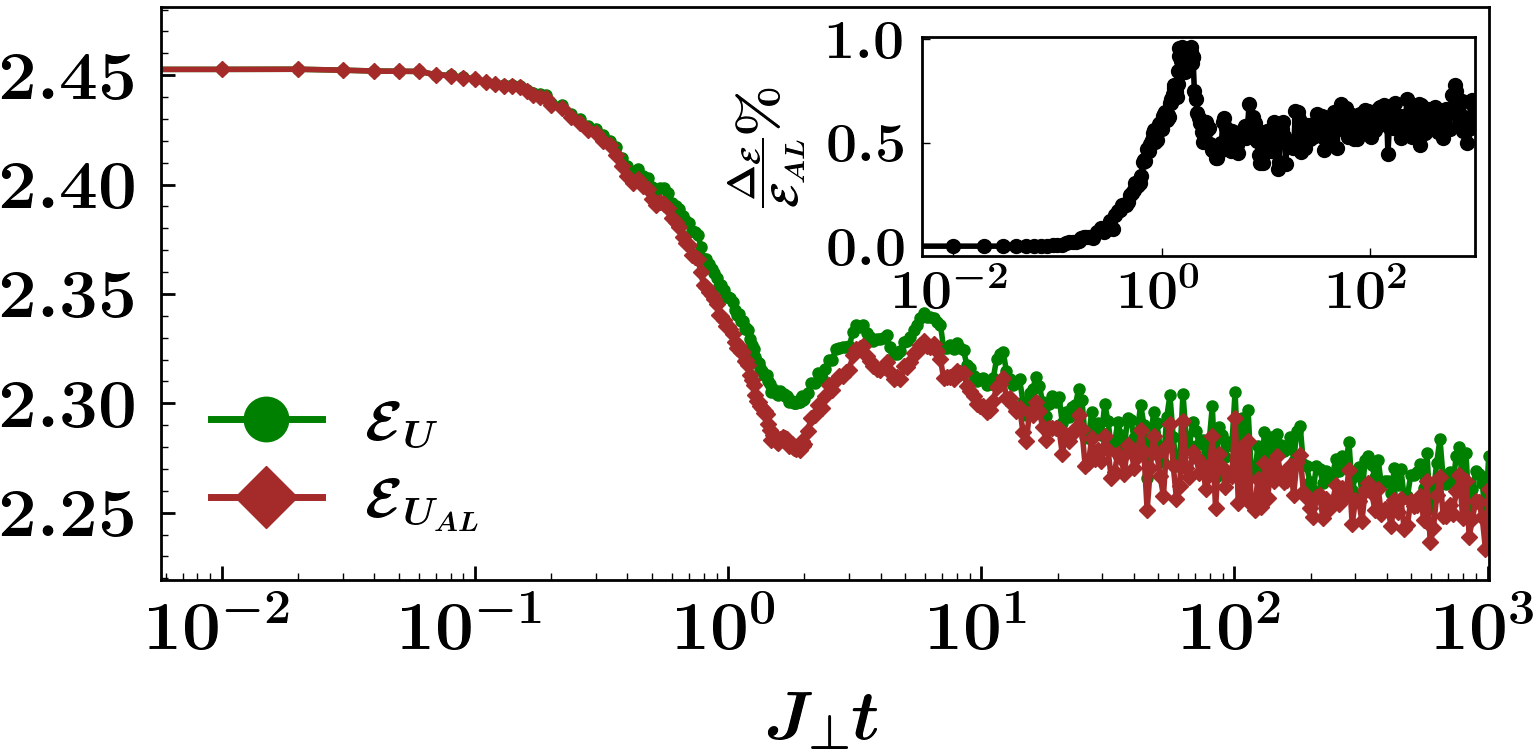}
    \put(-2.2, 50.6){\scalebox{1.5}{\textbf{(a)}}} 
  \end{overpic}
  \hspace{1.7cm}  
  \begin{overpic}[width=1\columnwidth]{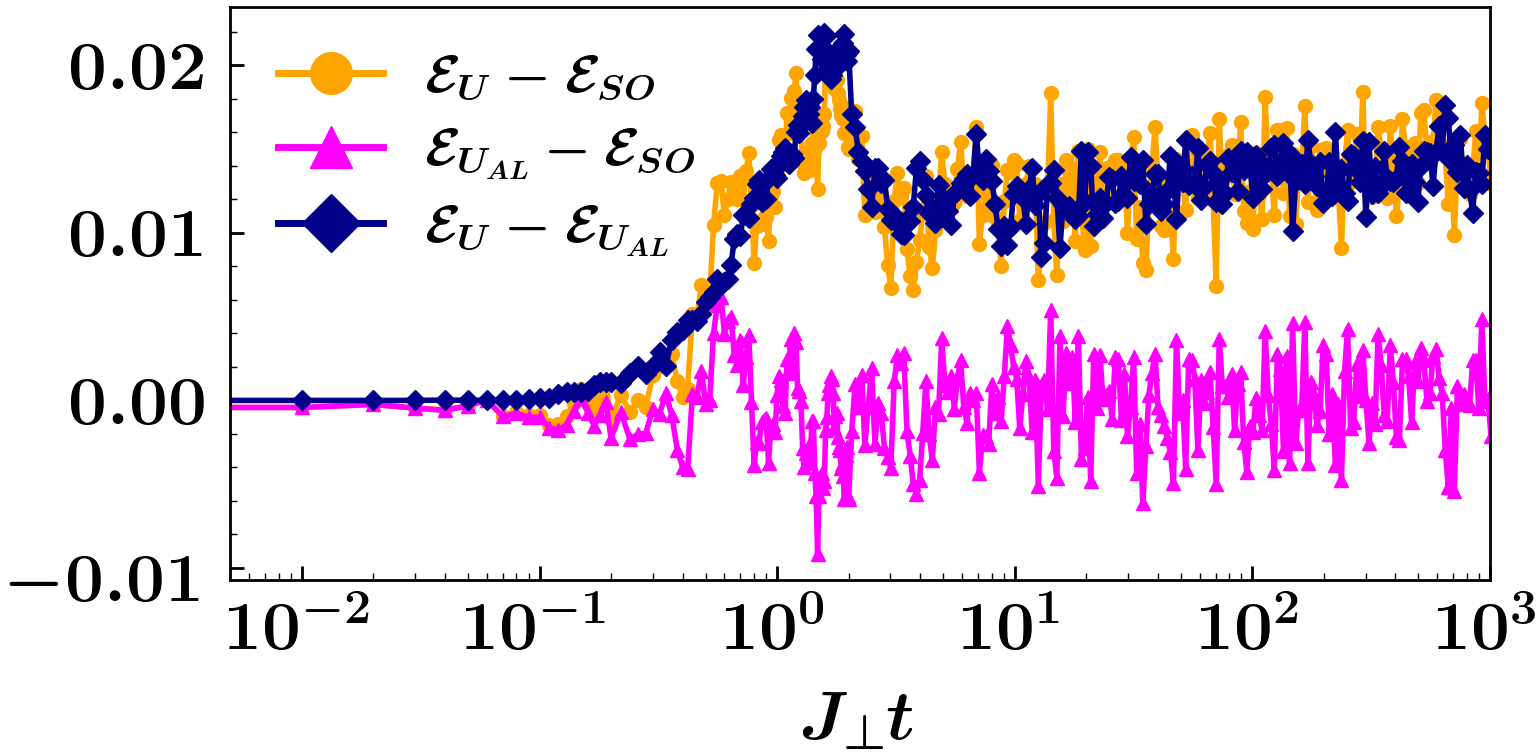}
   \put(-2.2, 50.6){\scalebox{1.5}{\textbf{(b)}}}  
  \end{overpic}
   \caption{\justifying (a) Local ergotropy as a function of time for two different unitary operators used to perform the numerical simulations: $\mathcal{E}_{U_{AL}}$ (brown rhombus) and $\mathcal{E}_{U}$ (green circle). The inset shows the relative difference between the two curves, $\Delta_{\mathcal{E}}=\mathcal{E}_{U}-\mathcal{E}_{U_{AL}}$, normalized to $\mathcal{E}_{U_{AL}}$, in percentage, as a function of time. (b) The behaviour in time of $\mathcal{E}_{U}-\mathcal{E}_{U_{AL}}$ (darkblue diamond), $\mathcal{E}_{U_{AL}}-\mathcal{E}_{SO}$ (fuchsia triangle) and $\mathcal{E}_{U}-\mathcal{E}_{SO}$ (orange circle).
  Results are obtained by averaging over $10^3$ disorder realizations, with chain length $N=8$ and the system in the MBL phase, for interaction strength $J_z/J_{\perp}=0.2$ and disorder strength $W/J_{\perp}=5.0$.} 
  \label{fig:differentU}
\end{figure}

In the panel (a) of Fig.\,\ref{fig:differentU}, we present the time evolution of local ergotropy for two different cases. In the first case, $\mathcal{E}_{U_{AL}}$ was computed in the numerical simulations using $\hat{U}_{AL}(t)$ as the unitary operator to extract work, without requiring the use of the BOA. In the second case, the most general unitary operator $\hat{U}(t)=\hat{U}_{AL}(t)\cdot \hat{U}_1(t)$, which incorporates BOA, was employed to compute $\mathcal{E}_{U}$. 

A comparison shows that the BOA increases the value of $\mathcal{E}_{U_{AL}}$ by approximately $1\%$. This can be understood in terms of the slowly increasing correlations in the XXZ model, which cause the optimal $\mathcal{E}_{U_{AL}}$ for the closed two-spin subsystem $S$ to be very close to the optimal value obtained when considering the environment $E$ (i.e., the remaining $N-2$ spins of the chain).

In the panel (b) of Fig.\,\ref{fig:differentU}, the discrepancy between the local ergotropy calculated using the BOA and the one without employing the BOA, $\mathcal{E}_{U}-\mathcal{E}_{U_{AL}}$, is presented, along with the differences between these two local ergotropies and the switch-off ergotropy: $\mathcal{E}_{U_{AL}}-\mathcal{E}_{SO}$ and $\mathcal{E}_{U}-\mathcal{E}_{SO}$. 
The investigation reveals that the work gained by utilising the BOA for local ergotropy is equivalent to the amount acquired from the local with respect to the switch-off ergotropy.
This outcome can be attributed to the fact that the unitary operator $\hat{U}_{AL}$ optimises exclusively within the $S$ subsystem Hilbert space, thereby precluding the utilisation of energy from the interactions with environment after the discharging gate is applied.

\section{Entangled initial state}\label{sec:entanglement_initial_state}
In the preceding sections of this work, we have used the Néel state $\ket{\psi}_N$ as the initial state for the dynamics due to its zero entanglement entropy and high mean energy (our charged state). Here, we present an investigation of the case where the initial state is not a product state but instead has a non-zero entanglement contribution, exploring whether entanglement can serve as a resource for extracting more work from the subsystem $S$.

\begin{figure}[htbp!]
  \includegraphics[width=1\columnwidth]{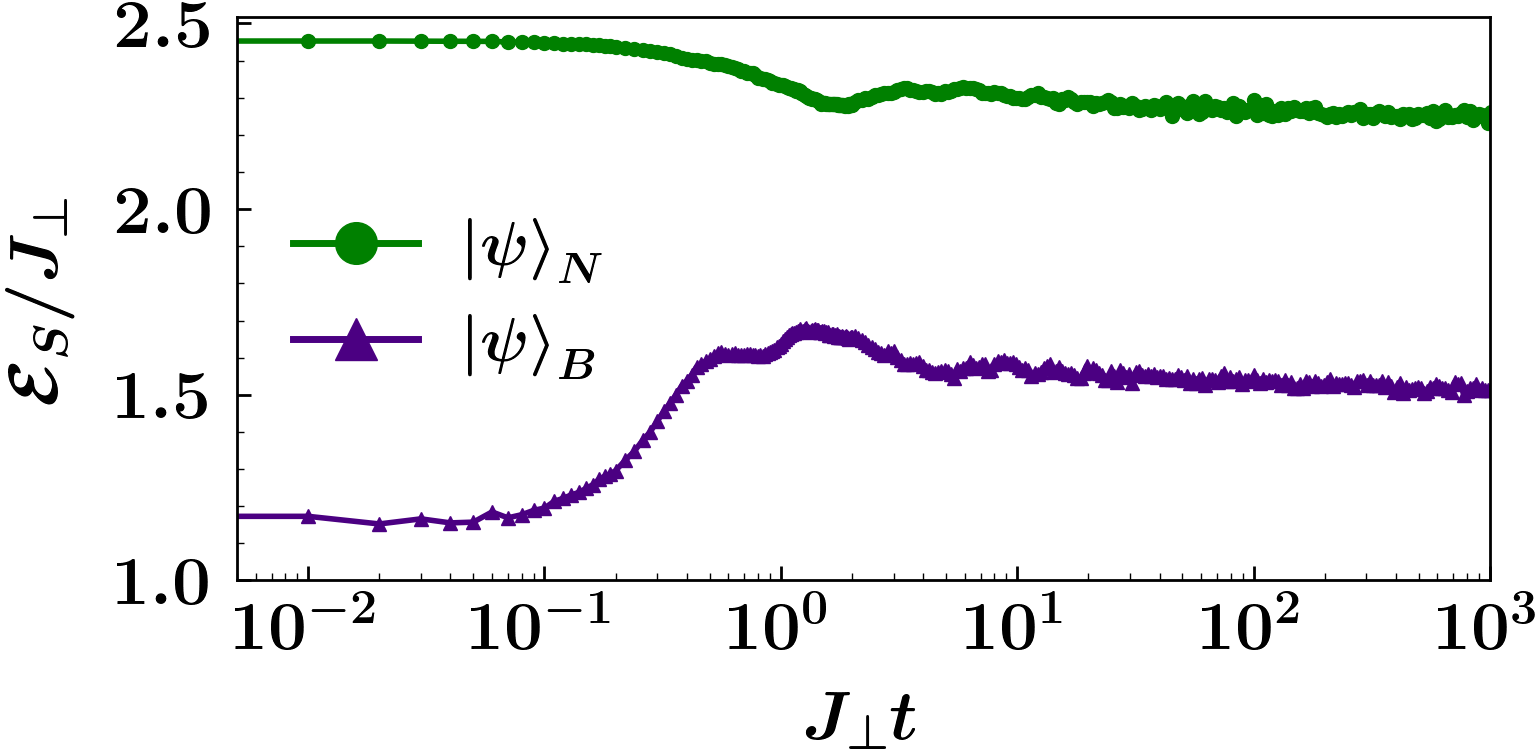}
\caption{\justifying Local ergotropy as a function of time in the MBL phase, for $J_z/J_{\perp}=0.2$ and $W/J_{\perp}=5$, for two different initial states: $\ket{\psi}_N$ (green circle), and $\ket{\psi}_B$ (indigo triangle). Results are obtained by averaging over $10^3$ disorder realizations, with chain length $N=8$, without using the BOA to compute the extractable work.}
  \label{fig:Bell}
\end{figure}

For the purpose of this investigation, we adopted a state in which both the first and last qubit of the chain do not exhibit entanglement with their neighbours. The remaining qubits were grouped into pairs of singlets. Thus, the initial state is 
\begin{equation}
    \ket{\psi}_B=\ket{\uparrow, \Psi^-,\Psi^-\dots,\Psi^-,\downarrow},
\end{equation}
where $\ket{\Psi^-}$ is the Bell state
\begin{equation}
    \ket{\Psi^-}=\frac{\ket{\uparrow}\otimes\ket{\downarrow}-\ket{\downarrow}\otimes\ket{\uparrow}}{\sqrt{2}}.
\end{equation}
Fig.\,\ref{fig:Bell} presents a comparative analysis of the local ergotropy behaviour over time in two cases: when the initial state is $\ket{\psi}_N$ or $\ket{\psi}_B$. After an initial transient period, which is nearly identical for both cases and during which the ergotropy trends differ, both exhibit a logarithmic decrease over time, as expected in the MBL phase. Besides the trend, the amount of extractable work is lower compared to the initial Néel state. This result is expected, as a greater portion of the system's energy is stored in interactions, making it unavailable for work extraction.

\bibliography{Bibliografia.bib}